\documentclass[twocolumn,showpacs,showkeys,preprintnumbers,superscriptaddress,amsmath,floatfix,amssymb,secnumarabic,nofootinbib]{revtex4}
\usepackage[colorlinks=true]{hyperref}
\usepackage{graphicx}
\usepackage{braket}
\usepackage{amsmath}
\usepackage{epsfig}
\usepackage{float}
\usepackage{color}

\usepackage{bm}

\def\({\left(}
\def\){\right)}
\def\[{\left[}
\def\]{\right]}

\newcommand{\diag}[1]{ {\rm diag} \, \left( #1 \right) }
\newcommand{\Tr}{ {\rm Tr} \, }

\newcommand{\beq} {\begin{eqnarray}}
\newcommand{\eeq} {\end{eqnarray}}

\newcommand{\comment}[1]{}

\begin{document}
\sloppy

\title{Path integral representation for the Hubbard model with reduced number of Lefschetz thimbles}

\author{M. V. Ulybyshev}
\email{ulybyshev.m@gmail.com}
        \affiliation{Institute for Theoretical and Experimental Physics (ITEP),117218 Russia, Moscow, B. Cheremushkinskaya str. 25}

\author{S. N. Valgushev}
\email{semen.valgushev@physik.uni-regensburg.de}
        \affiliation{Institute for Theoretical and Experimental Physics (ITEP),117218 Russia, Moscow, B. Cheremushkinskaya str. 25}
        \affiliation{Institut f\"ur Theoretische Physik, Universit\"at Regensburg, 93053 Regensburg, Germany}

\begin{abstract}
The concept of Lefschetz thimble decomposition is one of the most promising possible modifications of Quantum Monte Carlo (QMC) algorithms aimed at alleviating the sign problem which appears in many interesting physical situations, e.g. in the Hubbard model away from half filling. In this approach one utilizes the fact that the integral over real variables with an integrand containing a complex fluctuating phase is equivalent to the sum of integrals over special manifolds in complex space (``Lefschetz thimbles''), each of them having a fixed complex phase factor. Thus, the sign problem can be reduced if the resulting sum contains terms with only a few different phases. We explore the complexity of the sign problem for the few-site Hubbard model on square lattice combining a semi-analytical study of saddle points and thimbles in small lattices with several steps in Euclidean time  with results of test QMC calculations.  We check different variants of conventional Hubbard-Stratonovich  transformation based on Gaussian integrals. On the basis of our analysis we reveal a regime with minimal number of relevant thimbles in the vicinity of half filling. In this regime we found only two relevant thimbles for the few-site lattices studied in the paper. There is also indirect evidence of the existence of this regime in more realistic systems with large number of Euclidean time slices. In addition, we derive a new non-Gaussian representation of the interaction term, where the number of relevant Lefschetz is also reduced in comparison with conventional Gaussian Hubbard-Stratonovich  transformation. 

\end{abstract}

\pacs{71.10.Fd, 71.27.+a}
\keywords{Hubbard model, strongly correlated electrons, Morse theory, Lefschetz thimbles}

\maketitle

\section*{Introduction}
The Hubbard model has been the focus of intense theoretical and numerical research for decades since its introduction as a  model for strongly-correlated electrons in condensed matter physics \cite{HubbardBands1,HubbardBands2}. Those studies became especially important when it was realized that the physics of high-temperature superconductors can be approximately described by the two-dimensional  Hubbard model with finite chemical potential \cite{Hubbard_Cuprates_review, Hubbard_Cuprates_review1}.
However, despite large efforts made in this field \cite{HubbardBook}, there is still no comprehensive analytic or numerical method which can solve the Hubbard model for arbitrary parameters.

In this paper, we investigate improvements of the determinantal Quantum Monte Carlo (QMC) method which is one of the most common numerical techniques applied in studies of the Hubbard model. The main advantage of this method is that it does not involve any additional physical assumptions beyond those made to derive the interacting tight-binding Hamiltonian. Thus, QMC results are among the most reliable in the field. However, QMC often suffers from the sign problem, when a strongly fluctuating phase factor appears in the final integrals.  In particular, the most interesting case, that of the square lattice Hubbard model away from half filling, suffers from the sign problem which prevents a comprehensive study of the superconducting state.

The most recent approach to the sign problem in Monte Carlo simulations has its origin in the so-called Lefschetz thimble decomposition of the path integral \cite{Witten1,Witten2}, which can be considered as a multidimensional generalization of the stationary phase method. Namely, the integration contour of the partition function is deformed into complex domain such that original integral is represented as a sum of integrals over steepest descent manifolds (``Lefschetz thimbles'') originating from critical points of the action. Complex phases of integrands are constant on those manifolds, thus it is possible to use this property in order to solve or at least soften the sign problem. This approach was first proposed as a possible solutions of the sign problem in lattice Quantum Chromodynamics (QCD) at finite chemical potential \cite{Cristoforetti1,Cristoforetti1A} and investigated in a number of papers \cite{Cristoforetti3, DiRenzo1, Cristoforetti4, Kikukawa1,Tanizaki2, Alexandru1}. It was also employed for non-perturbative calculation of quantum corrections to mean field solutions in the Hubbard model \cite{Cristoforetti2}.

The naive Lefschetz thimbles approach requires the knowledge of all saddle points in the space of complex fields and involves integration over non-trivial manifolds, which is typically a very challenging problem. In order to overcome this difficulty it was recently proposed to construct a contour in complex space (the so-called Generalized thimble) which approximates Lefschetz thimbles without \textit{a priory} knowledge of saddle points and thimbles \cite{Alexandru2,Nishimura1,Alexandru4}, and thus reduce fluctuations of the complex phase. This can be achieved with the help of Holomorphic flow \cite{Alexandru2,Nishimura1,Tanizaki3} or using Machine Learning methods \cite{Mori1, Mori2, Alexandru4}.
With these numerical developments it became possible to simulate at least some simple models \cite{Alexandru3,Alexandru4} on relatively small lattices.

One should keep in mind that this is not a complete solution of the sign problem in every case, but a way to make the problem milder by suppressing fluctuations of the phase factor appearing under the integral. 

Despite these improvements of numerical algorithms, the number of Lefschetz thimbles needed to approach the true value of the initial integral remains one of the most important characteristics which quantifies the utility of these methods. The reason for that is twofold. First of all, the sign problem might return in the form of the sum over contributions from different Lefschetz thimbles where each term has its own complex phase. As was shown in \cite{Tanizaki}, already on the trivial example of the one-site Hubbard model, this problem can be very dramatic. On the other hand, for the algorithms mentioned above it is important to have as small number of contributing thimbles as possible because the underlying probability distribution appears to be strongly multimodal \cite{Fukuma1, Alexandru5} and the algorithm tends to sample the vicinity of some thimble and does not explore other thimbles.

In this paper we explore various variants of the path integral representation of the Hubbard model based on different representations of the four-fermionic interaction term. Some of them are based on a conventional Hubbard-Stratonovich transformation with Gaussian integrals and one is based on a newly developed integral representation which mimics the discrete transformation used in Blankenbecler-Scalapino-Sugar (BSS) QMC. We show that the number of Lefschetz thimbles and thus the complexity of the sign problem is very dependent on the particular representation. On the basis of our analysis we propose the regimes with substantially reduced number of relevant thimbles. These regimes are the most promising for the Generalized thimble algorithm.  Combining these approaches with recently developed algorithms \cite{Alexandru2, Alexandru3}, one can try to explore the phase diagram of the Hubbard model at higher chemical potential and lower temperature than was previously possible with existing QMC schemes.

The paper is organized as follows. In the first section, we give some basic definitions and make a brief review of the mathematical basis for existing determinantal QMC algorithms. We also give a short introduction to the Lefschetz thimbles method. In the next section we study the scaling of the number of relevant thimbles with increasing lattice size for the few-site Hubbard model in the case when conventional Gaussian Hubbard-Stratonovich (HS) transformation is used for the decomposition of the interaction term. Here we identify the regime with the minimal number of relevant thimbles.  In the third section we present results of test QMC calculations complementary to the findings presented in the previous section. The fourth section is devoted to the derivation of an alternative non-Gaussian path integral representation and its application to the simplest examples of the one-site Hubbard model and the few-site Hubbard model.

\section{\label{BasicDefs}Basic definitions}

\subsection{The model}

QMC algorithms usually deal with the path integral representation of the partition function 
\begin{equation}
\mathcal{Z}=\Tr e^{-\beta \hat H},
\label{part_func}
\end{equation}
and the corresponding thermodynamic averages of observables
\begin{equation}
\langle \hat O \rangle =\frac{1}{\mathcal{Z}} \Tr( \hat O e^{-\beta \hat H}).
\label{observable}
\end{equation}
Here $\hat H$ and $\hat O$ are the Hamiltonian and some observable respectively and $\beta$ is the inverse temperature. 
The Hamiltonian  $\hat H$ usually consists of two parts:
\begin{eqnarray}
\hat H &=& \hat H_{(2)} + \hat H_{(4)} = \nonumber
\\
=\sum_{x,y,\sigma,\sigma'} t_{xy\sigma\sigma'} {\hat c^\dag_{x\sigma}} \hat c_{y\sigma'} &+& \sum_{x,y,\sigma,\sigma'} U_{xy\sigma\sigma'} {\hat n_{x\sigma}} \hat n_{y\sigma'},
\label{full_hamiltonian}
\end{eqnarray}
where indexes $x$ and $y$ denote lattice sites, $\sigma, \sigma'=\uparrow,\downarrow$ correspond to spin index and $\hat n_{x\sigma} = {\hat c^\dag_{x\sigma}} \hat c_{x\sigma}$.
The first part contains only bilinear fermionic terms which includes a tight-binding part as well as the chemical potential. The second part contains four-fermionic terms describing electron-electron interaction. We have included in  (\ref{full_hamiltonian}) the most general interaction term, which can be referred to as the ``extended Hubbard model''. The Hubbard model itself includes only local on-site interaction $\sum_x U \hat{n_{x\uparrow}}\hat{n_{x\downarrow}}$.

The path integral representation of the partition function (\ref{part_func}) starts from the Trotter decomposition:
\begin{equation}
  \Tr e^{-\beta \hat H} \approx \Tr \left({ e^{-\delta \hat H_{(2)}} e^{-\delta \hat H_{(4)}}  e^{-\delta \hat H_{(2)}} e^{-\delta \hat H_{(4)}} ... } \right).
  \label{trotter}
\end{equation}
After the decomposition we have a product of $N_t$ exponentials, which constitutes the Euclidean time extent of the lattice. $\delta$ is the step in Euclidean time: $N_t \delta = \beta$.
To transform the trace to the path integral representation, one introduces Grassmann coherent states $|\xi \rangle$ and Grassmann variables $\xi$ for each creation and annihilation operator.
Further details of the construction of the path integral representation can be found in  \cite{BuividovichPolikarpov, ITEPRealistic, SmithVonSmekal}, where it was done for Hubbard-Coulomb model on hexagonal lattice. 

We would like to highlight one stage in this derivation which is important for our study of the sign problem. Namely, the four-fermionic term $\hat H_{(4)}$ in the full Hamiltonian (\ref{full_hamiltonian}) should be converted into a fermion bilinear. This step is essential, because for the bilinear terms in the exponent we have a simple set of relations which allows us to convert the multidimensional integral over Grassmann variables into the form convenient for a Monte Carlo scheme. 

 There are two different ways to convert the interaction term into bilinear form. The first scheme is based on discrete auxiliary variables \cite{Hirsh1,Hirsh2}. An example of such a transformation follows from the identity:
\begin{eqnarray}
\label{discrete_HS}
  e^{-\delta U \hat n_\uparrow \hat n_\downarrow} = \frac{1}{2} \sum_{\nu=\pm 1} e^{2 i \xi \nu ( \hat n_\uparrow + \hat n_\downarrow-1)-\frac{1}{2}\delta  U ( \hat n_\uparrow + \hat n_\downarrow-1) },
  \\
 \tan^2 \xi=\tanh(\frac{\delta U}{4}). \nonumber
\end{eqnarray}
Note that the exponents on the r.h.s. of this identity are purely imaginary for repulsive interactions $U>0$. One can also write a variant of this transformation leading to purely real exponents. This and similar representations are used in the Blankenbecler-Scalapino-Sugar (BSS) QMC algorithm which is widely applied to the physics of the Hubbard model \cite{FateMottHubbard,Pseudogap}.  
Another variant is based on the usual Gaussian HS transformation:
\begin{eqnarray}
\label{continuous_HS_imag}
  e^{-\frac{\delta}{2}\sum_{x,y} U_{x,y} \hat n_x \hat n_y} \cong \int D \phi_x e^{- \frac{1}{2\delta} \sum_{x,y} \phi_x U^{-1}_{xy} \phi_y} e^{i \sum_x \phi_x \hat n_x}, \\
   e^{\frac{\delta}{2}\sum_{x,y} U_{x,y} \hat n_x \hat n_y} \cong \int D \phi_x e^{- \frac{1}{2\delta} \sum_{x,y} \phi_x U^{-1}_{xy} \phi_y} e^{ \sum_x \phi_x \hat n_x}.
\label{continuous_HS_real}
\end{eqnarray}
It can be used in two variants leading to real (\ref{continuous_HS_real}) and complex (\ref{continuous_HS_imag}) exponents. This representation has an important advantage in that it also works for non-local interactions, so that we do not need to introduce a new auxiliary field for every pair of interacting electrons. Thus it was used, for instance, for the Hubbard-Coulomb model \cite{ITEPRealistic, BuividovichUlybyshevReview, Assaad:14:1,plasmons,Korner}. However, in the case of pure Hubbard model with only on-site interaction the number of discrete auxiliary fields in the first representation (\ref{discrete_HS}) is equal to the number of continuous fields in (\ref{continuous_HS_imag}) or (\ref{continuous_HS_real}). Thus, due to smaller configuration space, the discrete representation is more advantageous at least if the sign problem is absent. 

Now let's turn to the appearance of the sign problem. 
In special cases where some additional symmetries (e.g. the time-reversal symmetry \cite{Wu:05:1}) exist, the extended Hubbard model is accessible to QMC simulations. In particular, they are possible in the case of a bipartite lattice. Thus we are going to concentrate on the following Hamiltonian written on a bipartite lattice with only the on-site interaction term:
\begin{eqnarray}
\label{Hamiltonian_part}
  \hat H = -\kappa \sum_{\langle x,y\rangle, \sigma}  {\hat c^\dag_{x\sigma}} \hat c_{y\sigma} &+& U \sum_{x} {\hat n_{x\uparrow}} \hat n_{x\downarrow} - \nonumber  \\
  -  \left({\frac{U}{2} - \mu}\right) \sum_x   ({\hat n_{x\uparrow}} &+& \hat n_{x\downarrow}-1).  
\end{eqnarray}
The tight-binding part includes only hopping to nearest neighbors. The chemical potential $\mu$ defines the shift from half-filling, which corresponds to $\mu=0.0$ in our notation. 
QMC algorithms in ideal situation (in the absence of the sign problem) need at least a semi-positive weight for auxiliary fields.
The bipartite lattice provides us with this possibility at half-filling, after a well-known trick which transforms spin-up and spin-down electrons ($\hat{c}_{x, \uparrow}$ and $\hat{c}_{x, \downarrow}$) to electrons and holes ($\hat{a}_x$ and $\hat{b}_x$):
\begin{equation}
\label{ParticleHole}
 \left\{ { {\hat{c}_{x, \uparrow}, \hat{c}^{\dagger}_{x, \uparrow} \to \hat{a}_x, \hat{a}^{\dagger}_x, } \atop
{\hat{c}_{x, \downarrow}, \hat{c}^{\dagger}_{x, \downarrow} \to \pm \hat{b}^{\dagger}_x, \pm \hat{b}_x} } \right. ,
\end{equation}
where the sign in the second line alternates depending on the sublattice.
The Hamiltonian  (\ref{Hamiltonian_part}) acquires the following form after the transition to the new variables :
\begin{eqnarray}
  \label{Hamiltonian_el_hol}
  \hat H = -\kappa \sum_{\langle x,y\rangle} (  \hat a^\dag_{x} \hat a_{y} + \hat b^\dag_{x} \hat b_{y} ) &+& \frac{U}{2} \sum_{x} (\hat n_{x, el.} - \hat n_{x, h.})^2 + \nonumber \\
  + \mu  \sum_x (\hat n_{x, el.} - \hat n_{x, h.}),
\end{eqnarray}
where $\hat n_{x, el.} = {\hat a^\dag_{x}} \hat a_{x}$ and $\hat n_{x, h.} = {\hat b^\dag_{x}} \hat b_{x}$ are the particle number operators for electrons and holes respectively.

Now we should make either the discrete (\ref{discrete_HS}) or the continuous (eq. (\ref{continuous_HS_imag}) and (\ref{continuous_HS_real}) ) transformation for each exponent in the expression (\ref{trotter}) where the interaction part of the full Hamiltonian appears. Thus, auxiliary fields acquire the Euclidean time index $t$ in addition to the spatial lattice site index $x$. Since the interaction is local, only one auxiliary field variable will appear per lattice site in both cases. 
In the case of the discrete transformation (\ref{discrete_HS}), we arrive at the following representation of the partition function (\ref{part_func}) as a sum over all possible values of $\nu_{x,t}$:
\begin{eqnarray}
  \mathcal{Z}_d = \sum_{\nu_{x,t}} \det D_{el.}(\nu_{x.t}) \det D_{h.}(\nu_{x,t}),
  \label{Z_discrete}
\end{eqnarray}
where $D_{el.}$ and $D_{h.}$ are fermionic operators for electrons and holes respectively:
\begin{eqnarray}
 D_{el.}(\nu_{x,t}) = I +\prod^{N_t}_{t=1} \left({ e^{-\delta (h+\mu)} \diag{ e^{2i \xi \nu_{x,t}} } }\right), \nonumber \\
 D_{h.}(\nu_{x,t}) = I +\prod^{N_t}_{t=1} \left({ e^{-\delta (h-\mu)} \diag{ e^{-2i \xi \nu_{x,t}} } }\right). 
  \label{M_discrete}
\end{eqnarray}
Both fermionic operators are $N_s \times N_s$ matrices where $N_s$ is the number of lattice sites in space, $h$ is the matrix of single-particle Hamiltonian which defines the tight-binding part in the expression (\ref{Hamiltonian_el_hol}).  The diagonal  $N_s \times N_s$ matrix ${\diag{ e^{-2i \xi \nu_{x,t}} }}$ includes all exponents with auxiliary fields belonging to a given Euclidean time slice $t$.

In the case of continuous auxiliary fields, we will write the HS transformation in more general way employing both real (\ref{continuous_HS_real}) and complex (\ref{continuous_HS_imag}) exponents:
\begin{eqnarray}
\frac{U}{2}( \hat n_{el.} - \hat n_{h.})^2 = \frac {\alpha U}{2}( \hat n_{el.} - \hat n_{h.})^2 - \nonumber \\
- \frac{(1-\alpha) U}{2} ( \hat n_{el.} + \hat n_{h.})^2 + (1-\alpha) U ( \hat n_{el.} + \hat n_{h.}).
\end{eqnarray}
Parameter $\alpha \in [0,1]$ defines the balance between real and complex exponents in the integral. 
The first four-fermionic term can be transformed into bilinear using (\ref{continuous_HS_imag}) and the second using (\ref{continuous_HS_real}). This is not the most general possible decomposition of four-fermionic terms into bilinear ones, but the most commonly used in QMC algorithms with continuous auxiliary fields. This representation was first proposed in \cite{complex1} and was also used in the recent paper \cite{Assaad_complex}. 
The partition function can be written as the following integral:
\begin{eqnarray}
  \mathcal{Z}_c = \int \mathcal{D} \phi_{x,t} \chi_{x,t} e^ {-S_{\alpha}}  \det M_{el.} \det M_{h.}, \nonumber \\
   S_\alpha(\phi_{x,t},\chi_{x,t}) = \sum_{x,t}  \frac  {\phi_{x,t}^2} {2 \alpha \delta U}  + \sum_{x,t}  \frac {(\chi_{x,t}- (1-\alpha) \delta U)^2} {2 (1-\alpha) \delta U},
  \label{Z_continuous}
\end{eqnarray}
where fermionic operators for continuous auxiliary fields are written as
\begin{eqnarray}
 M_{el.} = I +\prod^{N_t}_{t=1} \left({ e^{-\delta \left(h+\mu\right)} \diag{ e^{i \phi_{x,t}+\chi_{x,t}} } }\right), \nonumber \\
 M_{h.} = I +\prod^{N_t}_{t=1} \left({ e^{-\delta \left(h-\mu\right)} \diag{ e^{-i \phi_{x,t}+\chi_{x,t}} } }\right). 
  \label{M_continuous}
\end{eqnarray}
In subsequent derivations we will deal with the full action which includes both quadratic form and the logarithms of the fermionic determinants:
\begin{eqnarray}
S=S_{\alpha} - \ln (\det M_{el.} \det M_{h.}).
 \label{action_continuous}
\end{eqnarray}

In all cases, we disregard constant multipliers in the integrals since they are cancelled in the computation of observables (\ref{observable}). An important point is that both representations of the partition function reproduce the exact result only in the limit $\delta \rightarrow 0$ due to approximations introduced by the Trotter decomposition (\ref{trotter}).

One can easily see that the fermionic determinants for electrons and holes both in the discrete and the continuous cases are complex conjugated to each other at half filling. This means that we can safely use the expressions under the sum  in (\ref{Z_discrete}) or the integral in (\ref{Z_continuous}) as the weight for sampling auxiliary fields. Away from half-filling, we should employ ``reweighting'' where auxiliary fields are sampled according to the modulus of the corresponding expressions in (\ref{Z_discrete}) or (\ref{Z_continuous}) and the remaining complex phase factor is included in observable. However, this procedure suffers from cancellation of terms with opposite phases and thus its domain of applicability is limited to rather low values of the chemical potential and rather high temperatures. 

\begin{figure}
        \centering
        \includegraphics[scale=0.3]{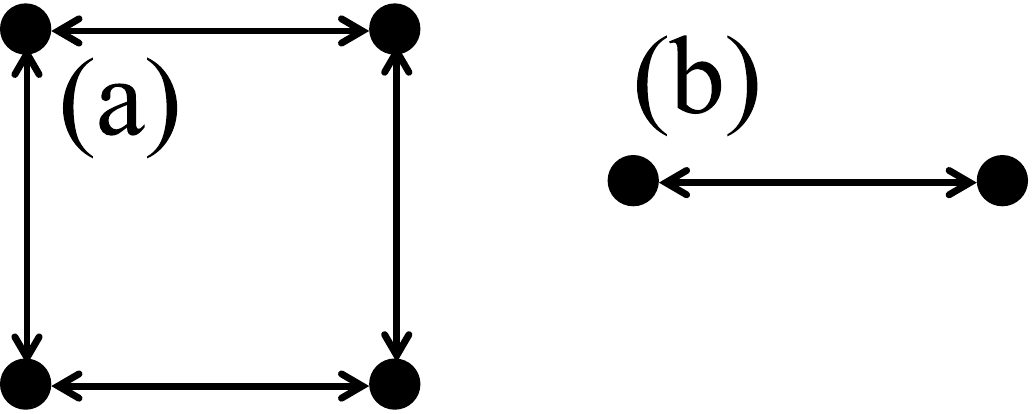}
        \caption{Lattices studied in the paper: (a) 4-site and (b) two-site. The arrows denote hoppings between lattice sites.}
        \label{fig:lattices}
\end{figure}

Despite being less effective for the pure Hubbard model with only on-site interaction, the second approach has one important advantage away from half-filling. Since it is based on continuous variables, the integration manifold can be shifted from real space to complex space, leading to the ``complexification'' of the auxiliary fields. If no singular points are crossed during this shift, Cauchy's integral theorem guarantees the same answer. The alternative integration manifold in the complex plane can be chosen in a special way to reduce the fluctuations of the phase of the integrand in eq. (\ref{Z_continuous}). This is the main idea of the Generalized thimbles algorithm. 

Since the finding of Lefschetz thimbles in many-dimensional complex space is a non-trivial numerical task, in the current paper we will explicitly consider only the small lattices with $N_s=1,2,4$  and $N_t=1,2,3,4$. Corresponding square lattices with two and four sites and periodical boundary conditions are shown in the figure \ref{fig:lattices}. The hopping amplitude $\kappa$ can be any complex number, only hoppings to nearest-neighbours are taken into account. The single-particle Hamiltonian takes the form 
\begin{eqnarray}
 h=\begin{pmatrix} 0 & -\kappa \\ -\bar \kappa & -0   \end{pmatrix}
 \label{2site_hamiltonian}
\end{eqnarray}
when $N_s=2$ and
\begin{eqnarray}
 h=\begin{pmatrix} 0 & -\kappa & -\kappa & 0 \\ -\bar \kappa & 0 & 0 & -\kappa \\ -\bar \kappa & 0 & 0 & -\kappa \\ 0 & -\bar \kappa & -\bar \kappa & 0    \end{pmatrix}.
 \label{4site_hamiltonian}
\end{eqnarray}
when $N_s=4$. 
For the one-site Hubbard model the Hamiltonian $h$ can obviously be disregarded. One should also note that the case of $N_t=1$ is exact for the one-site Hubbard model as there is no error associated with the discretization of Euclidean time. Indeed, there are no alternating exponents in the Trotter decomposition (\ref{trotter}) if the tight-binding part is absent. Thus it is sufficient to have only one step in Euclidean time. It automatically guarantees that the path integral  representation (\ref{Z_discrete}) or (\ref{Z_continuous}) exactly reproduces the initial partition function (\ref{part_func}).

Despite the very limited system sizes under consideration, all approaches we discuss are still fully applicable for real computations on lattices with large $N_s$ and $N_t$.

\subsection{Lefschetz thimbles method}

In order to illustrate the basic ideas of the Lefschetz thimbles method, we start from the most general form of integrals appearing in QMC with continuous auxiliary fields:
\begin{equation}
\label{S_general}
\mathcal{Z}\left(\beta, \mu, \dots \right) = \int_{\mathbb{R}^N} d^N x e^{-S(\beta, \mu, \dots, x)}.
\end{equation}
If we consider continuation of this integral in domain of complex-valued  variables $x \in \mathbb{C}^N$, then due to Cauchy theorem one can choose any appropriate contour in complex space for integration. A representation with particularly useful properties can be constructed with the help of Morse theory (or Picard-Lefschetz theory) and is known as Lefschetz thimble decomposition of the path integral \cite{Witten1,Witten2}:
\begin{eqnarray}
\label{thimbles_sum}
\mathcal{Z}\left(\beta, \mu, \dots \right) &=& \sum_\sigma k_\sigma\left(\beta, \mu, \dots \right) \mathcal{Z_\sigma}\left(\beta, \mu, \dots \right),\\
\label{thimble_integral}
\mathcal{Z_\sigma}\left(\beta, \mu, \dots \right) &=& \int_{\mathcal{I}_\sigma\left(\beta, \mu, \dots \right)} d^N x e^{-S(\beta, \mu, \dots, x)},
\end{eqnarray}
where $\sigma$ labels all complex saddle points $z_\sigma \left(\beta, \mu, \dots \right) \in \mathbb{C}$ of the action:
\begin{equation}
\label{saddle}
{\left.{ \frac{\partial S} {\partial x} } \right| }_{x=z_\sigma\left(\beta, \mu, \dots \right)} = 0,
\end{equation}
integer-valued coefficients $k_\sigma\left(\beta, \mu, \dots \right)$ are so-called intersection numbers and $\mathcal{I_\sigma}\left(\beta, \mu, \dots \right)$ are steepest descent (Lefschetz thimble) manifolds. Here we have emphasized dependence of all important quantities and objects on parameters for clarity and will omit this in what follows. This relation is valid if saddle points are non-degenerate and isolated (for generalization in the case of gauge theory see \cite{Witten1}). Degenerate saddle points can appear due to spontaneous breaking of some continuous symmetry, and in this case, the symmetry should be explicitly broken by some small term in Hamiltonian and all results should be extrapolated to the limit where the symmetry is restored. 

In order to construct the Lefschetz thimble manifold $\mathcal{I}_\sigma$ corresponding to a given complex saddle point $z_\sigma$ we use the gradient flow equation:
\begin{equation}
\label{flow}
\frac{dx}{d\tau}=\overline{ \frac{\partial S}{\partial x}},
\end{equation}
with the following boundary conditions:
\begin{equation}
\label{thimble}
x\in\mathcal{I}_\sigma:x(\tau)=x, x(\tau\rightarrow -\infty) \rightarrow z_\sigma.
\end{equation}
This equation defines the evolution of the complex variable $x$ with respect to the fictitious flow time $\tau$, and all such solutions constitute the thimble manifold.

Analogously, we define another important type of manifold, the so-called anti-thimble $\mathcal{K}_\sigma$ which consist of all possible solution of the flow equations (\ref{flow}) which end up at a given saddle point $z_\sigma$:
\begin{equation}
\label{anti-thimble}
x\in\mathcal{K}_\sigma:x(\tau)=x, x(\tau\rightarrow + \infty) \rightarrow z_\sigma. 
\end{equation}
With the help of anti-thimbles one can compute integer-valued coefficients $k_\sigma$ in the expression (\ref{thimbles_sum}) by counting the number of intersection of $\mathcal{K}_\sigma$ with original integration contour $\mathbb{R}^N$:
\begin{equation}
\label{intersection_number}
k_\sigma = \langle \mathcal{K}_\sigma, \mathbb{R}^N \rangle.
\end{equation}

Both thimbles and anti-thimbles are $N$-dimensional real manifolds in $\mathbb{C}^N$. Two basic properties which make them useful are the following. First of all, the real part of the action $\mbox{Re}\, S$ monotonically increases along the thimble and monotonically decreases along the anti-thimble if we start from the saddle point. Secondly, the imaginary part of the action $\mbox{Im}\, S$ stays constant along both of them. It follows that neither thimbles nor anti-thimbles can intersect each other, neither of saddle points can be connected by some thimble in a general situation (with a very important exception which is discussed below) and all integrals on the r.h.s. of the expression (\ref{thimbles_sum}) are convergent.

It is due to constant complex phases on thimbles this method became attractive for studying the sign problem in QMC simulations. There is nevertheless a residual sign problem due to non-trivial complex-valued volume element on the thimble which is however soft and can be overcome. In practice, thimbles can be constructed using their tangent spaces in the vicinity of saddle points. Namely, at each saddle point we can compute $2N \times 2N$ matrix of the second derivatives of $\mbox{Re}\, S$ over real and imaginary part of complex variable $x$. This matrix has exactly $N$ positive and $N$ negative eigenvalues. The corresponding eigenvectors define the tangent space for the thimble and the anti-thimble respectively and provide us with initial conditions for the flow equations.

The main subtlety in this theory is a Stokes phenomenon which happens when at some values of parameters (so-called Stokes rays) there exist two or more distinct saddle points connected by some thimble. This can only happen when the imaginary part of actions at these saddle points coincide: $\mbox{Im}\, S(z_\sigma) = \mbox{Im}\, S(z_{\sigma^\prime})$. Then thimble integrals $\mathcal{Z}_\sigma$ associated with these saddle points exhibit jumps which should be compensated by the jump in coefficients $k_\sigma$ in order to ensure validity of Cauchy theorem. Jumps in intersection numbers appear due to change in the structure of anti-thimbles. Consequently, some coefficients $k_\sigma$ might become zero or non-zero and the structure of the sum (\ref{thimbles_sum}) can change dramatically, thus any reasonable QMC algorithm based on the thimble decomposition should correctly account for them. This makes direct application of the thimble decomposition very impractical, however the very existence of such decomposition motivates development of algorithms which will approximate thimbles in some automatic manner and minimize sign problem, like those mentioned in the introduction.

The general sign problem generated by the fluctuating phase in (\ref{S_general}) is substituted by the sign problem generated by different phase factors appeared in the sum over thimbles (\ref{thimbles_sum}):
\begin{eqnarray}
\label{thimbles_sum_with_phases}
\mathcal{Z} = \sum_\sigma k_\sigma e^{-i\, \mathrm{Im}\, S(z_\sigma)} \int_{\mathcal{I}_\sigma} d^N x e^{-\mathrm{Re} \, S(x)},
\end{eqnarray}
where we write down complex factors associated with different saddle points  explicitly. We say that thimble is ``relevant'' if it has a nonzero intersection number $k_\sigma$ and thus participates in this sum. The number of relevant thimbles, their weight, and the distribution of imaginary part of action at corresponding (relevant) saddle points define the remaining complexity of the sign problem. The smaller the number of relevant saddle points, the less severe the sign problem in (\ref{thimbles_sum}). An ideal situation is of course when we have just one relevant thimble or if only one thimble is important in the sum (\ref{thimbles_sum}) due to dominating absolute value of integral over it.

\subsection{Hybrid Monte Carlo and problems with ergodicity}

Hybrid Monte Carlo (HMC) algorithm is now the most widely used technique to update continuous fields during the Markov process in QMC \cite{deTar}. Details of this method impose some limitations on the possible path integral representations, so we give the brief description of the method. In HMC we employ artificial dynamics to make the updates of continuous fields. The main steps in  the algorithm can be described as follows:
\begin{itemize}
    \item Artificial momentum $\theta_{x,t}$ is introduced for each continuous auxiliary field $\psi_{x,t}$.
    \item The classical Hamiltonian for the artificial evolution is written as $H=1/2 \sum_{x,t} \theta^2_{x,t} + S(\psi_{x,t})$, where the action  $S(\psi_{x,t})$ includes both quadratic form and logarithms  of fermionic determinants (see eq. (\ref{action_continuous})).
    \item The update of both auxiliary fields and momenta is performed through the solution of classical dynamics equations according to this Hamiltonian. The Metropolis accept-reject step is made in the end of trajectory. 
\end{itemize}

Hamiltonian updates used in HMC impose important limitation on the ergodicity of the method. Namely, these updates can not penetrate through the manifolds formed by configurations with zero fermionic determinant, because the action $S(\psi_{x,t})$ goes to infinity at these configurations. If the dimensionality of these manifolds is equal to $N-1$ within the general $N$-dimensional integration manifold, then we have ``domain walls'' and the single HMC process can do only integration within the single region surrounded by domain walls. In order to penetrate though the domain walls we need some other, non-Hamiltonian updates. Below we will show that this situation indeed emerges in some particular cases for the path integrals for the Hubbard model. 

If $\alpha=0$ the complex exponents and the auxiliary fields $\phi(x,t)$ disappear from the integral (\ref{Z_continuous}) and fermionic determinants both for electrons and holes are purely real functions. They are identical at half-filling and start to differ at nonzero $\mu$ thus the fluctuating sign (but not complex phase) appears in the integrand in (\ref{Z_continuous}). Since all functions are real, all relevant saddle points and thimbles are also within the real subspace $\mathbb{R}^N$. It means that we are automatically within the representation of the partition function through the sum over thimbles (\ref{thimbles_sum}) even without any shift to the complex plane: the real subspace is simply divided between thimbles attached to relevant real saddles. Thus we do not even need to search for some manifolds in complex space. Away from half-filling some of these saddles have positive sign and some of them have negative sign.   Simple counting of degrees of freedom shows that the manifold of the zero points of determinant $\det M_{el.} \det M_{h.}$ has dimensionality $N-1$ in $\mathbb{R}^N$ in general case. It means that the regions belonging to different thimbles in $\mathbb{R}^N$ are separated by ``domain walls'' of configurations with zero probability and HMC can not  explore the full phase space, even at half filling. 
This phenomenon was already observed in \cite{Assaad_complex,Cristoforetti2}, where the representation with only real exponents was used for QMC calculations.
This is not an issue if we already know in advance the dominant saddle(s) and want to compute integrals over corresponding thimbles as it was done in \cite{Cristoforetti2}, but ideally one should construct some numerical procedure which doesn't need \textit{a priori} knowledge. 

The same situation with ``domain walls'' emerges also in the opposite limit where $\alpha=1$  and real exponents completely disappear from the integral  (\ref{Z_continuous}). In this case, we can make the Hubbard-Stratonovich transformation coupling the auxiliary field with spin degrees of freedom and compare the final integral with the one derived in terms of electrons and holes (\ref{M_continuous}). If there are only hoppings between sublattices (no mass term and no chemical potential is introduced), the following relation for the fermionic operator in (\ref{M_continuous}) can be proved for $\alpha=1$:
\begin{equation}
  \det M_{el.} (\phi_{x,t}) =  F (\phi_{x,t}) e^{i \sum_{x,t}\phi_{x,t} }.
  \label{M_zero_modes_relation}
\end{equation}
The function $F(\phi_{x,t})$ is real and it is not equal to the sum of squares, it can change the sign. The overall product of determinants
\begin{equation}
\det M_{el.}  \det M_{h.} = {F (\phi_{x,t})}^2
  \label{M_square_cont}
\end{equation}
is again equal to the square of some real function. It means that if the zero points of determinant exist (if they are not eliminated by, e.g. some explicit mass term in one-particle Hamiltonian), their manifold again has dimensionality $N-1$ in $\mathbb{R}^N$.
This fact can be noticed already in the simplest case of the lattice with $N_s=2$ and $N_t=1$. The  fermionic determinant for this lattice takes the form
\begin{eqnarray}
    \det M_{el.}  \det M_{h.} = e^{-2\beta \kappa} \times \nonumber \\
   \left({ (1+e^{2\beta \kappa}) \cos(\frac{\phi_1-\phi_2}{2}) + 2 e^{2\beta \kappa} \cos(\frac{\phi_1+\phi_2}{2})   }\right)^2
\end{eqnarray}
in the limit $\alpha=1$.

We see that both limits of purely complex ($\alpha=1$) and purely real ($\alpha=0$) exponents are not entirely suitable for the HMC simulations due to the ``domain walls'' within the integration domain. Here we should stress again that insertion of explicit mass term in one-particle Hamiltonian as it was made in previous calculations on hexagonal lattice \cite{ITEPRealistic,SmithVonSmekal, vacancies} completely eliminates the problems with ergodicity since configurations with zero fermionic determinant are absent \cite{BuividovichPolikarpov}. The price is that the mass term can introduce bias towards some particular channel of spontaneous symmetry breaking. 
Thus, if we want to make calculations without this bias, some intermediate value of $\alpha$ should be used. First,  we analyze both cases of $\alpha=1$ and $\alpha=0$ in order to give a comprehensive picture of the sign problem. Then we study  intermediate values of $\alpha$ where the situation smoothly evolves between these too limits.

\section{\label{sec:Gauss}Lefschetz thimbles and Gaussian Hubbard-Stratonovich transformation}

\begin{figure}
        \centering
        \includegraphics[scale=0.7]{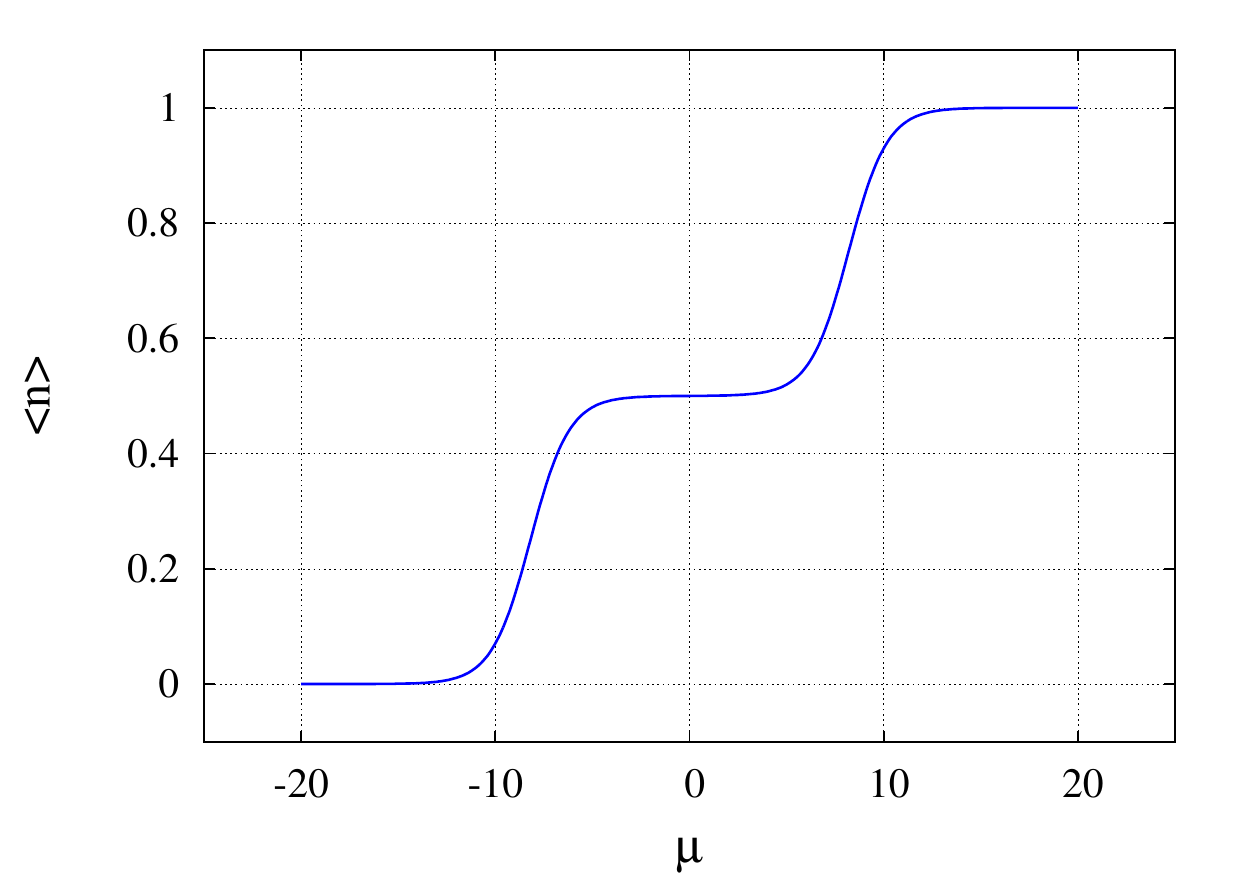}
        \caption{Average number of particles $\langle \hat n \rangle=\langle \hat a^{\dag} \hat a \rangle $ for the one-site Hubbard model. $U\beta=15.0$.}
        \label{fig:n}
%thimbles_gauss_full.nb
\end{figure}

\begin{figure*}
        \centering
        \includegraphics[scale=0.7]{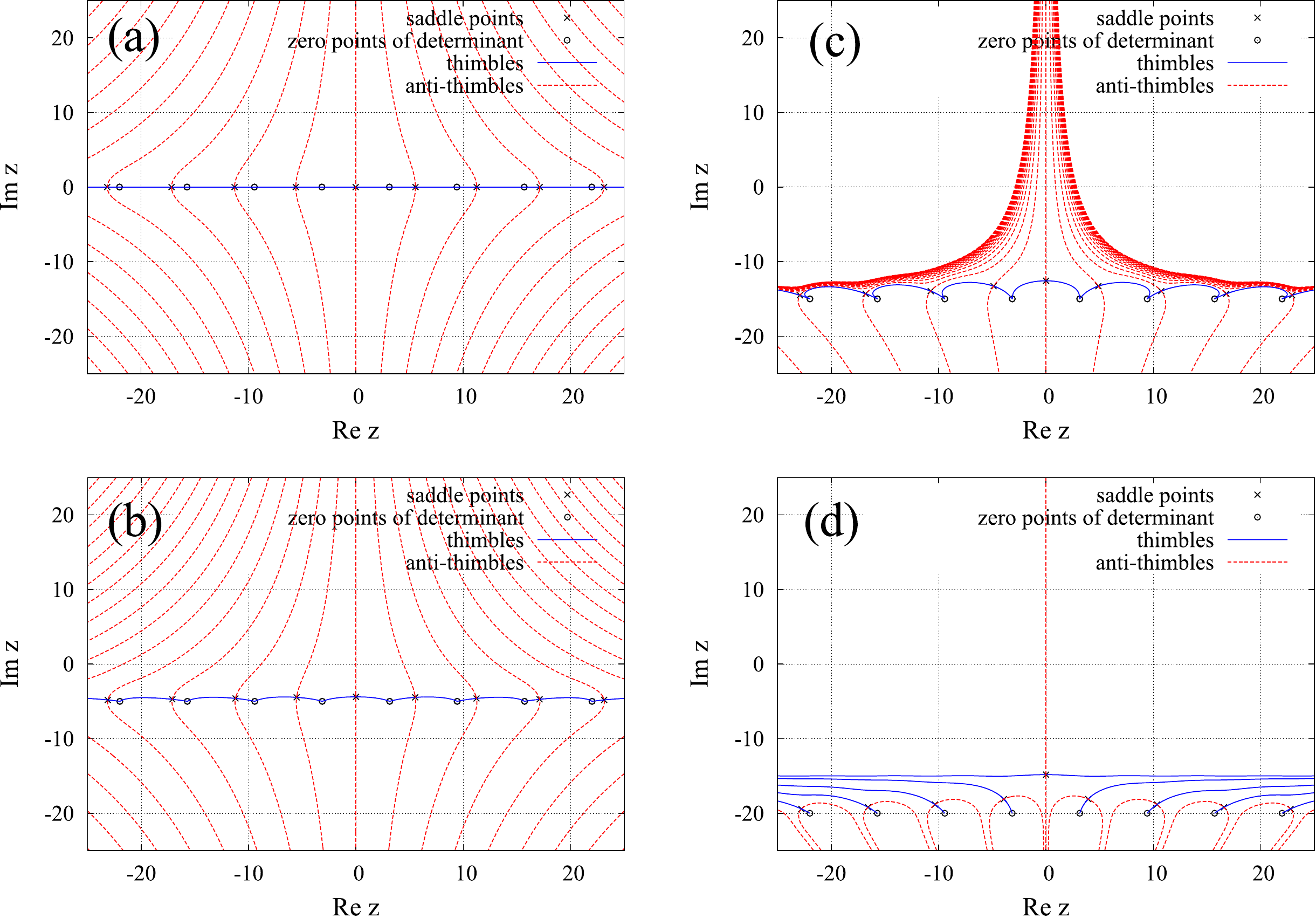}
        \caption{Thimbles and anti-thimbles for one-site Hubbard model in the Gaussian representation at various values of chemical potential.The action is written in (\ref{action_one_site}),  $U\beta=15.0$. (a)  Half filling ($\mu=0$).  The real axis is divided by ``zeros'' of fermionic determinant into infinite number of thimbles. Corresponding anti-thimbles end up at infinity $\mbox{Im}\, z \rightarrow \pm \infty$. (b) $\beta \mu=5.0$. The number of relevant thimbles is still infinite but all relevant saddles are shifted in the complex plane from the real axis. (c) $\beta \mu=15.0$. There is still infinite number of relevant saddles, but the Stokes phenomenon is very close to appearance. (d) $\beta \mu=20.0$. The Stokes phenomenon is occurred. Only one relevant thimble remained.}
        \label{fig:one_site_gauss_thimbles}
%thimbles_gauss_full.nb
\end{figure*}

Now we are going to explore how the Lefschetz thimbles approach works for different variants of Gaussian HS transformation. 
In order to estimate the complexity of the sign problem, we estimate the number or relevant thimbles, calculate their phases and estimate their weight in the sum (\ref{thimbles_sum}). We use built-in routine \textit{FindRoot} from \textit{Mathematica} in order to find saddle points and routine \textit{NDSolve} in order to solve the flow equations (\ref{flow}). In this exploratory study we restrict ourselves to quite small lattice sizes because we use explicit expressions for fermionic determinants in computations. After finding the saddle points we estimate the absolute value of the integrals over thimbles in order to identify their real contribution to the overall sum (\ref{thimbles_sum}). We base on the first approximation for the action in the vicinity of saddle points:
\begin{eqnarray}
    S \approx S|_{x_0} + \left. { \frac{1}{2} {\frac{\partial S } {\partial x^i \partial x^j }}} \right|_{x_0} (x^i-x^i_0) (x^j-x^j_0).
    \label{S_expansion}
\end{eqnarray}
Thus the integral over thimble can be estimated as the Gaussian one and the weight of thimble is defined by $\exp{(-W)}$, where
\begin{eqnarray}
    W=\mbox{Re} S|_{x_0} + \frac{1}{2} \log \det D_2.
    \label{W}
\end{eqnarray}
$D_2$ is the matrix of the second derivatives of the real part of the action calculated over coordinates within the thimble (denoted as $t_i$) in the vicinity of saddle point:
\begin{eqnarray}
D_2 = \left. {\frac{\partial \mbox{Re}S } {\partial t^i \partial t^j }}\right|_{t=t(x_0)}.
\label{D_2}
\end{eqnarray}
For real saddles it coincides with the matrix of the second derivatives of the action within $\mathbb{R}^N$. For complex saddles we calculate the $2N \times 2N $ matrix $\mathcal{D}_2$ of the second derivatives of $\mbox{Re}S$ over real and imaginary parts of complex variable $x$ and compute the $\log \det D_2$ as the sum of logarithms of positive eigenvalues of this matrix.

\subsection{Gaussian HS transformation with only complex exponents}

Following \cite{Tanizaki}, we start from the one-site Hubbard model because it allows to illustrate some basic concepts by plotting thimbles and anti-thimbles in simple 2D figures. According to definitions in Section (\ref{BasicDefs}), the action in the path integral representation for the partition function (\ref{Z_continuous}) of this model can be written as:
\begin{equation}
\label{action_one_site}
 S(x) = \frac{x^2}{2 \beta U} -\ln \left( { (1+ e^{i x -\beta \mu} ) (1+ e^{-i x + \beta \mu})} \right).
\end{equation}
We used $\alpha=0$ thus only complex exponents are left in the action. The model is exactly solvable: at low temperatures ($\beta U \gg 1$) there is sharp transition in the number of particles $\langle \hat n \rangle =\langle \hat a^\dag \hat a \rangle $ when the absolute value of chemical potential becomes comparable to the interaction strength $U$. The number of particles as a function of chemical potential is plotted in the figure \ref{fig:n}.

\begin{figure}
        \centering
        \includegraphics[scale=1.0, trim=-0cm 0 0 0]{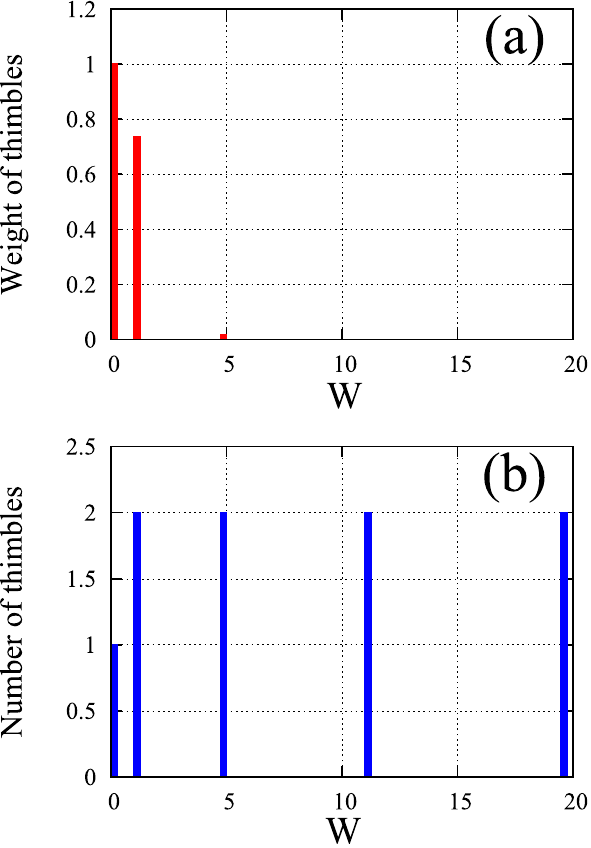}
        \caption{Weighted (a)  and normal (b) histogram showing the relative importance of relevant saddles for the one-site Hubbard model in the Gaussian representation at half-filling ($\mu=0$). The action is written in (\ref{action_one_site}), $U\beta=15.0$. Weight of thimbles is counted with respect to the vacuum one.}
        \label{fig:one_site_gauss_weight}
%thimbles_gauss_full.nb
\end{figure}

\begin{figure}
        \centering
        \includegraphics[scale=0.7]{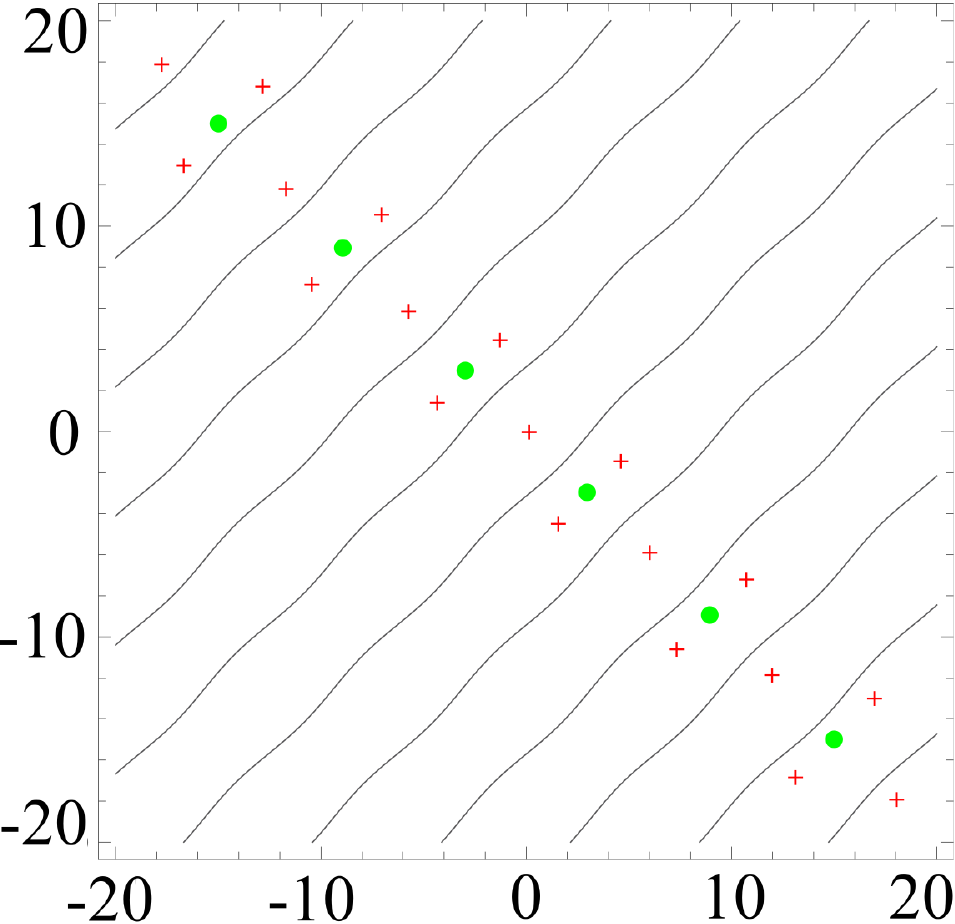}
        \caption{Position of the real saddles at half-filling for the two-site Hubbard model with one time slice in the Gaussian representation with only complex exponents. 
        Relevant (positive) saddles are marked with red crosses and irrelevant (negative)  saddles are marked by green circles. The lines represent the configurations of the fields $\phi_i$ where fermionic determinant is equal to zero. 
        The action is written according to eq. (\ref{Z_continuous}), (\ref{M_continuous}), (\ref{2site_hamiltonian}) with parameters: $U\beta=15.0$, $\kappa\beta=3.0$,  $\alpha=1$.}
        \label{fig:2D_gauss_positions}
%saddles_half_filling_2D_Nt1_gauss.nb
\end{figure}

\begin{figure*}
        \centering
        \includegraphics[scale=1.1, trim=-0cm 0 0 0]{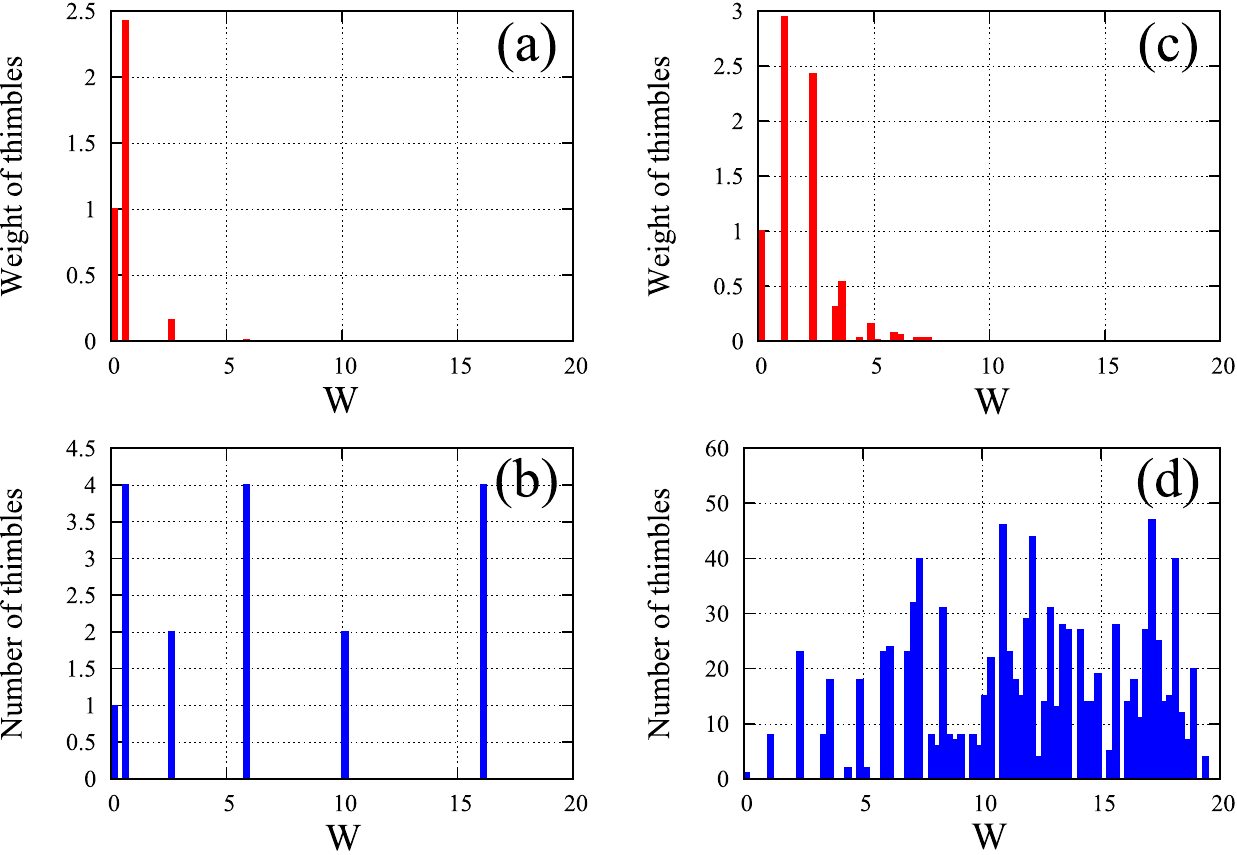}
        \caption{Weighted and normal histogram for relevant thimbles at half-filling for two-site lattice ((a) and (b)) and for four-site lattice ((c) and (d)). Gaussian representation with complex exponents is used, $N_t=1$ in both cases.
        The action is written according to equations (\ref{Z_continuous}), (\ref{M_continuous}), (\ref{2site_hamiltonian}) with parameters: $U\beta=15.0$, $\kappa\beta=3.0$, $\alpha=1$. Weight of thimbles is shown with respect to the vacuum one.}
        \label{fig:2D_4D_gauss_weight}
%saddles_half_filling_2D_Nt1_gauss.nb
\end{figure*}

Saddle points, thimbles and anti-thimbles for this model are shown in the figure \ref{fig:one_site_gauss_thimbles} for four different situations. The first case (fig. \ref{fig:one_site_gauss_thimbles}\textcolor{red}{a}) corresponds to half-filling ($\mu=0$); the second case (fig. \ref{fig:one_site_gauss_thimbles}\textcolor{red}{b}) corresponds to intermediate chemical potential ($\mu=U/3$) and the last two plots (fig. \ref{fig:one_site_gauss_thimbles}\textcolor{red}{c} and \ref{fig:one_site_gauss_thimbles}\textcolor{red}{d})  correspond to the case of large chemical potential ($\mu \geqslant U$) which is comparable to the interaction strength and causes the transition in the average number of particles $\langle \hat n \rangle$. These figures illustrate the key properties of thimbles and anti-thimbles which are important for further consideration. Both thimbles and anti-thimbles start from saddle points. Since the real part of the action monotonically increases along thimbles, they can end up either at infinity or at the points where the fermionic determinant is equal to zero, because $\mbox{Re}\, S$ tends to infinity in both cases. Anti-thimbles should end up in the region where $\mbox{Re}\, S$ monotonically decreases. In this model it corresponds to some direction at infinity. We will show further that there are also other possibilities.

At small and intermediate chemical potential ($\mu<U$) there is an infinite number of anti-thimbles crossing the real axis. Thus, there are an infinitely large amount of relevant saddle points which should be included into  the sum (\ref{thimbles_sum}). The relative importance of the different terms in the sum (\ref{thimbles_sum}) was estimated for this model in \cite{Tanizaki} within the saddle points approximation, where the whole integral over the thimble is substituted by the value of the exponent at the corresponding saddle point $e^{-S(z_\sigma)}$. The zeroth saddle at $x=0$ is of course dominant but one should take into account $\approx 5$ thimbles to reach reasonable precision at intermediate chemical potential around the transition point. 
This hierarchy is illustrated in the figure \ref{fig:one_site_gauss_weight} using the approximations described in eq. (\ref{S_expansion}-\ref{D_2}). This is the typical plot which we will use for the estimation of the relative importance of thimbles in various situations. The lower plot is the histogram showing the number of thimbles which have their values of weight $W_\sigma$ (see eq. (\ref{W})) within the given interval with respect to the thimble with the largest weight $W_0$. The upper plot is the ``weighted'' histogram. It means that the height of each bar increases by the relative weight $\exp\left({- (W_\sigma-W_0)}\right)$ of the thimble with respect to the vacuum one if $W_\sigma$ of the thimble belongs to the given interval. The weighted histogram (fig. \ref{fig:one_site_gauss_weight}\textcolor{red}{a}) clearly shows that the ``vacuum'' saddle at zero $x$ still dominates. The weight of all further thimbles (there are two of them contributing to each bar, these thimbles are symmetrical with respect to $x=0$) rapidly decreases with increased distance from the vacuum $x=0$.

\begin{figure*}
        \centering
        \includegraphics[scale=1.7]{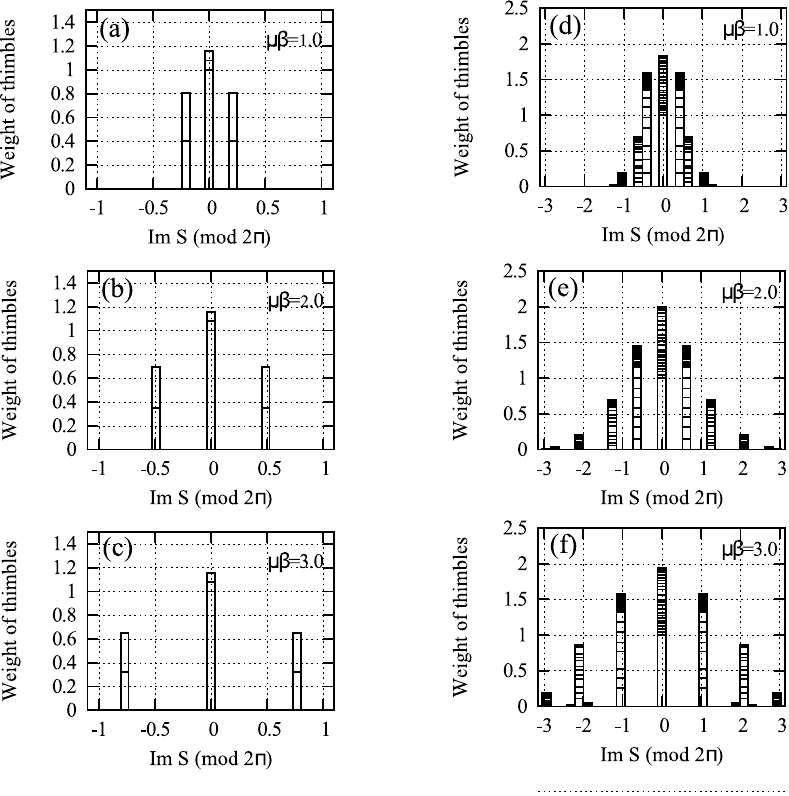}
        \caption{Stacked histograms showing the evolution of real saddles which were relevant at half-filling when we increase the chemical potential. In general, they acquire growing complex phases. Each bar shows the weights of several thimbles whose $\mbox{Im}\, S$ lie within the given interval. Contributions of these thimbles are separated by horizontal lines within the bars. The weight of former ``vacuum'' saddle is taken as unity.  The calculation is made for two-site lattice ((a), (b) and (c)) and four-site lattice ((d), (e) and (f)). $N_t=1$ in both cases. The action is written according to equations (\ref{Z_continuous}), (\ref{M_continuous}), (\ref{2site_hamiltonian}),(\ref{4site_hamiltonian}) with parameters: $U\beta=15.0$, $\kappa\beta=3.0$, $\alpha=1$, $\mu\beta=1.0,\,2.0,\,3.0$. } 
       \label{fig:2D_4D_gauss_weight_finite_mu}
\end{figure*}

%TODO explain stacked histograms

\begin{figure*}
        \centering
        \includegraphics[scale=1.1, trim=-0cm 0 0 0]{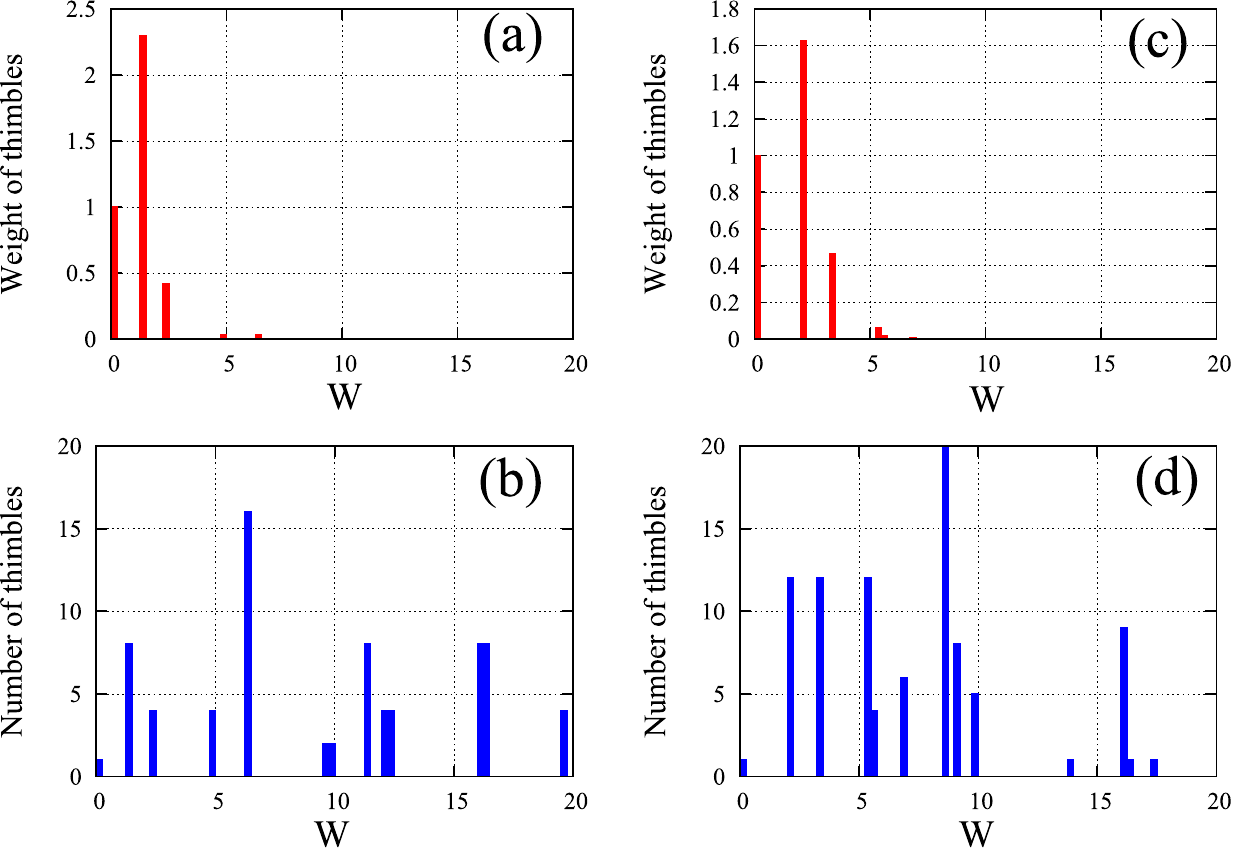}
        \caption{Weighted and normal histograms for relevant thimbles at half-filling for two-site lattice with two ((a) and (b)) and three ((c) and (d)) Euclidean time slices. Gaussian representation with only complex exponents is used, the action is written according to equations (\ref{Z_continuous}), (\ref{M_continuous}), (\ref{2site_hamiltonian}) with parameters: $U\beta=15.0$, $\kappa\beta=3.0$, $\alpha=1$. Weight of thimbles is shown with respect to the vacuum one.}
        \label{fig:2D_Nt2_Nt3_gauss_weight}
%saddles_half_filling_2D_Nt1_gauss.nb
\end{figure*}

The main question is how this situation scales when the overall lattice size $N = N_s N_t$ increases. A full derivation of the exact scaling law for the number of relevant thimbles is probably unfeasible in the general case. Thus, our task is to find, empirically, whether the number of important saddles increases with increasing lattice size. 
We will study the region $\mu<U$, since  the chemical potential is usually smaller than the typical scale of the on-site interaction in reality. For instance, in graphene, new physical phenomena emerge if the chemical potential crosses the van Hove singularity \cite{Lifshitz} which is of the order of the hopping (2.7 eV), while on-site interaction is of the order of 10 eV \cite{Wehling}. 
We will consider the two-site Hubbard model on the lattice with $N_t=1,2,3$ and the four-site Hubbard model with $N_t=1$. Action is constructed according to (\ref{Z_continuous}) and (\ref{M_continuous}) with the single-particle Hamiltonian defined in (\ref{2site_hamiltonian}) and (\ref{4site_hamiltonian}) and $\alpha=1$. The general form of the action in these cases can be written as
\begin{equation}
\label{general_action}
S(\phi)=\sum_i \frac{\phi_i^2}{2\delta U} - \ln \left({\det M_{el.}(\phi) \det M_{h.} (\phi)}\right),
\end{equation}
where $\delta=\beta / N_t$.
At half-filling all relevant saddles are obviously located within the real subspace $\mathbb{R}^N$ and the same is true for all relevant thimbles. It means that the tangent subspace for the anti-thimbles is oriented perpendicular to $\mathbb{R}^N$ in the vicinity of these relevant saddles. 
It also means that once we introduce nonzero $\mu$ and the former real saddles shift from $\mathbb{R}^N$ into complex space, their intersection number still remains equal to 1 if the shift is not that large. 
Moreover, we can expect that additional complex saddles (which in principle might become relevant) do not play important role in the overall sum (\ref{thimbles_sum}) for small chemical potential, especially  if we do not pass trough a phase transition. Within all these approximations, we can estimate the complexity of the sign problem at relatively small chemical potential by looking at the distribution of real saddles at half filling and then tracing the shift of former real saddles to the complex plane at finite $\mu$.

The figure \ref{fig:2D_gauss_positions} illustrates the position of real saddle points and configurations with zero fermionic determinant in the two-site Hubbard model with $N_t=1$ and $\alpha=1$ at half-filling for $U\beta=15.0$ and $\kappa \beta = 3.0$. 
Now, two types of saddle points appear at half filling within the real subspace. The classification is made using the matrix of the second derivatives $D_2$ calculated entirely within the real subspace. ``Positive'' and ``negative'' saddles have positive- and negative-defined matrix $D_2$ respectively. Only positive saddles are relevant, because the thimbles corresponding to the saddle points with negative-defined matrix $D_2$ cross the real subspace only at a set of points with dimension less then $N$.

This feature can be easily understood in the model with the double-well potential $(\phi^2-m^2)^2$. There are two stable minima at $\phi=\pm m$. These points correspond to relevant saddle points, while unstable equilibrium at $\phi=0$ corresponds to an irrelevant saddle point.  Nevertheless, it plays important role in the geometry of thimbles: this point separates two thimbles which start from two stable minima, while its own thimble is perpendicular to the real axis.
This property can be generalized to the $N$-dimensional case. 
If some saddle has its matrix $\mathcal{D}_2$ with $M \leq N$ negative eigenvalues, this saddle is irrelevant. But it still has $N-M$ positive eigenvalues. It means that the intersection of the thimble emanating from this saddle with real subspace $\mathbb{R}^N$ has dimension $N-M$. This set of points forms a domain wall between two thimbles within the real subspace if $M=1$ or just the manifold of sunk points for relevant thimbles if $M>1$. 

To sum it all up, only ``positive'' real saddles are relevant when we are looking at a system at half filling. According to the figure \ref{fig:2D_gauss_positions}, there is again an infinite set of relevant saddle points. The thimbles are separated by both the domain walls formed by configurations with zero fermionic determinant and by the lines originating from the ``negative'' (irrelevant) real saddles. Unlike the domain walls, these lines are penetrable for the HMC updates. In general, the appearance of the infinite number of relevant thimbles is the consequence of the periodicity of the fermionic determinant (\ref{M_continuous}) as a function of auxiliary fields in $\mathbb{R}^N$. 

The figure \ref{fig:2D_4D_gauss_weight} shows the weight of thimbles at half filling for two-site and four-site lattice with $N_t=1$. A comparison with the same plots for the one-site  model \ref{fig:one_site_gauss_weight} shows that the situation rapidly becomes worse. The ``vacuum'' saddle at $\phi_1=\phi_2=0$ is no more dominant. Instead of it the spatially nonuniform saddles play the main role in the sum and their number increases non-linearly with increasing $N_s$.  
For example, the number of thimbles needed to be taken into account for four-site lattice already approaches $100$.

\begin{figure*}
        \centering
        \includegraphics[scale=1.1, trim=-0cm 0 0 0]{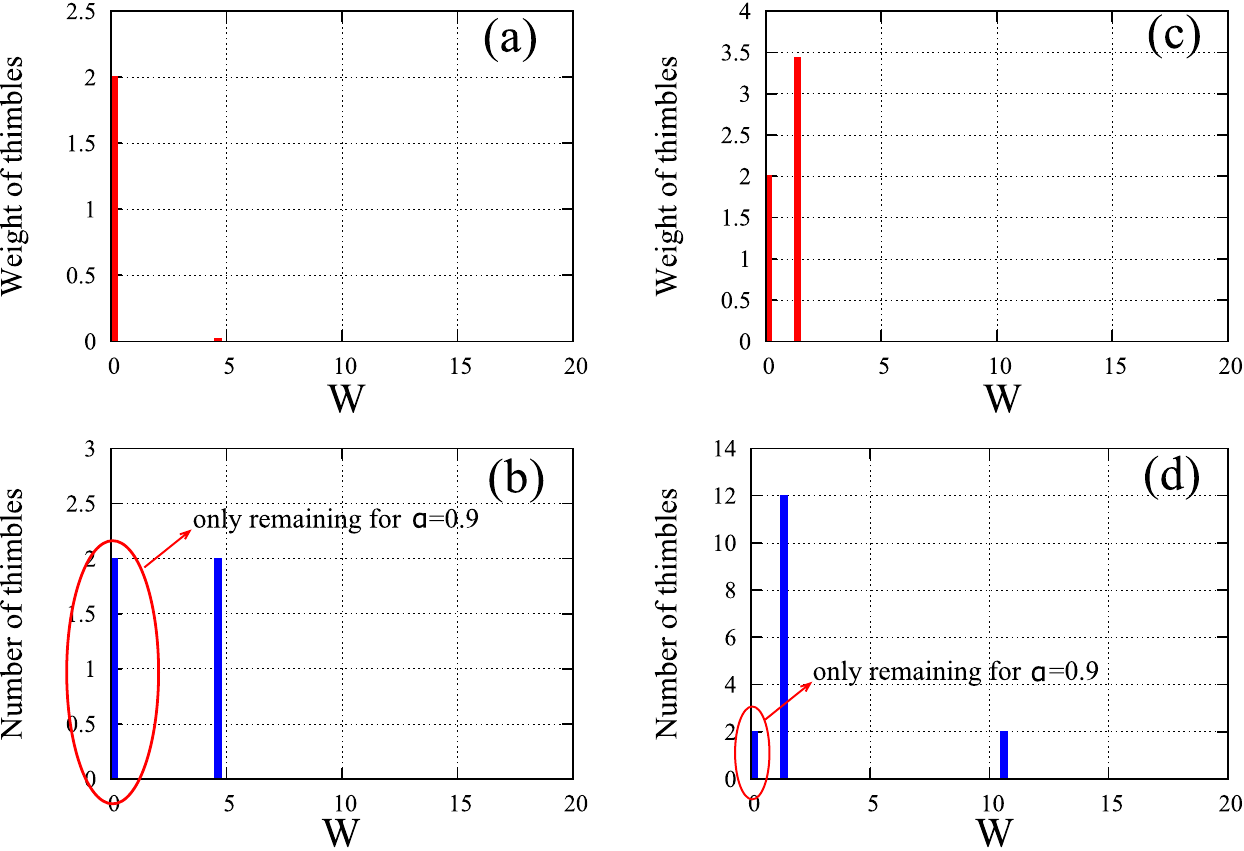}
        \caption{Weighted and normal histograms for relevant thimbles at half-filling for two-site ((a) and (b)) and four-site ((c) and (d)) lattice. Gaussian representation with only real exponents is used, $N_t=1$ in all cases.  The action is written according to equations (\ref{Z_continuous}), (\ref{M_continuous}), (\ref{2site_hamiltonian}) and (\ref{4site_hamiltonian}) with parameters: $U\beta=15.0$, $\kappa\beta=3.0$, $\alpha=0$. Weight of thimbles is shown with respect to the vacuum one. We also mark the thimbles, which remain relevant if the mixed representation is used with $\alpha=0.9$.}
        \label{fig:2D_4D_6D_gauss_real_weight}
%saddles_half_filling_2D_Nt1_gauss.nb
\end{figure*}

Once we go away from half-filling, former real saddles move to the complex plane and acquire complex phases. This process is illustrated in the figure \ref{fig:2D_4D_gauss_weight_finite_mu} for lattices with $N_S=2,4$ and $N_t=1$. These stacked histograms show the distribution of the imaginary part of the action for former real saddles which presumable remain relevant for relatively small chemical potential. The  delimiters inside  each bar mark the share of one thimble within the bar. One can clearly see how the complexity of the sign problem increases with increased number of thimbles: the distribution of $\mbox{Im} S$ becomes much broader and denser for the four-site model. It is directly connected with the appearance of new types of non-uniform saddles which are not equivalent to each other due to translational or rotational symmetry  (equivalent saddles have the same $\mbox{Im} S$). Once the lattice size is increased these saddles fill the unit circle in complex plane more and more densely thus increasing the sign problem.

The cases of the two-site model with $N_t=2$ and $N_t=3$ are shown in the figure \ref{fig:2D_Nt2_Nt3_gauss_weight}. The number of non-uniform saddles increases again, but in the continuous limit the situation might actually become much better. 
The reason is that almost all of these non-vacuum saddle points are lying already behind the ``domain wall'': the situation from the figure \ref{fig:2D_gauss_positions} is reproduced also for larger lattices. Thus these thimbles appear in the sum only due to the presence of the ``domain walls''. If $N_t$ is increased, the coefficient before the squared fields in the action  increases (see eq. (\ref{Z_continuous})) thus the configurations with large values of the fields $\phi_{x,t}$ are suppressed. On the other hand, the overall scale for the distance between ``domain walls'' is fixed by the period $2\pi$ of the fermionic determinant and this period is independent on $N_t$. Thus the configurations behind the domain wall with  $\phi_{x,t}>\pi$ are effectively suppressed and we a left with very limited number of thimbles which are really important.
This phenomena can be already seen in the figures \ref{fig:2D_Nt2_Nt3_gauss_weight}\textcolor{red}{a} and \ref{fig:2D_Nt2_Nt3_gauss_weight}\textcolor{red}{c}), where the weight of the second bar (the lowest ``non-vacuum''  thimbles) decreases which increased $N_t$.

\begin{figure*}
        \centering
        \includegraphics[scale=0.9, trim=-0cm 0 0 0]{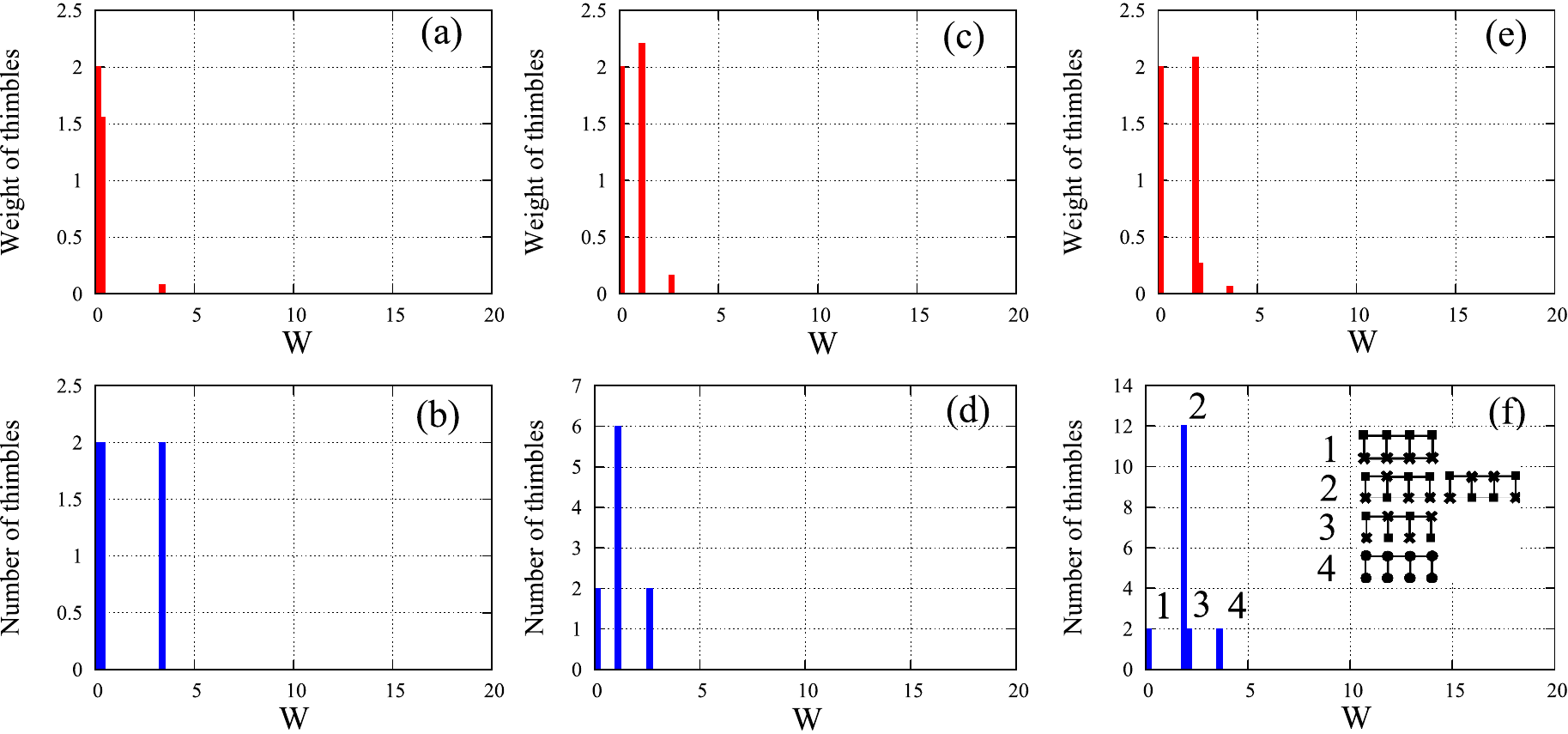}
        \caption{Weighted and normal histograms for relevant thimbles at half-filling for two-site lattice with two ((a) and (b)), three ((c) and (d)), and four ((e) and (f)) Euclidean time slices. Gaussian representation with only real exponents is used, the action is written according to equations (\ref{Z_continuous}), (\ref{M_continuous}), (\ref{2site_hamiltonian}) with parameters: $U\beta=15.0$, $\kappa\beta=3.0$, $\alpha=0$. Weight of thimbles is shown with respect to the vacuum one. Inset in the figure (f) shows schematic pictures of relevant saddle points corresponding to each of the four bars in the histogram. }
        \label{fig:2D_Nt2_Nt3_gauss_real_weight}
%saddles_half_filling_2D_Nt1_gauss.nb
\end{figure*}

\begin{figure*}
        \centering
        \includegraphics[scale=0.9, trim=-0cm 0 0 0]{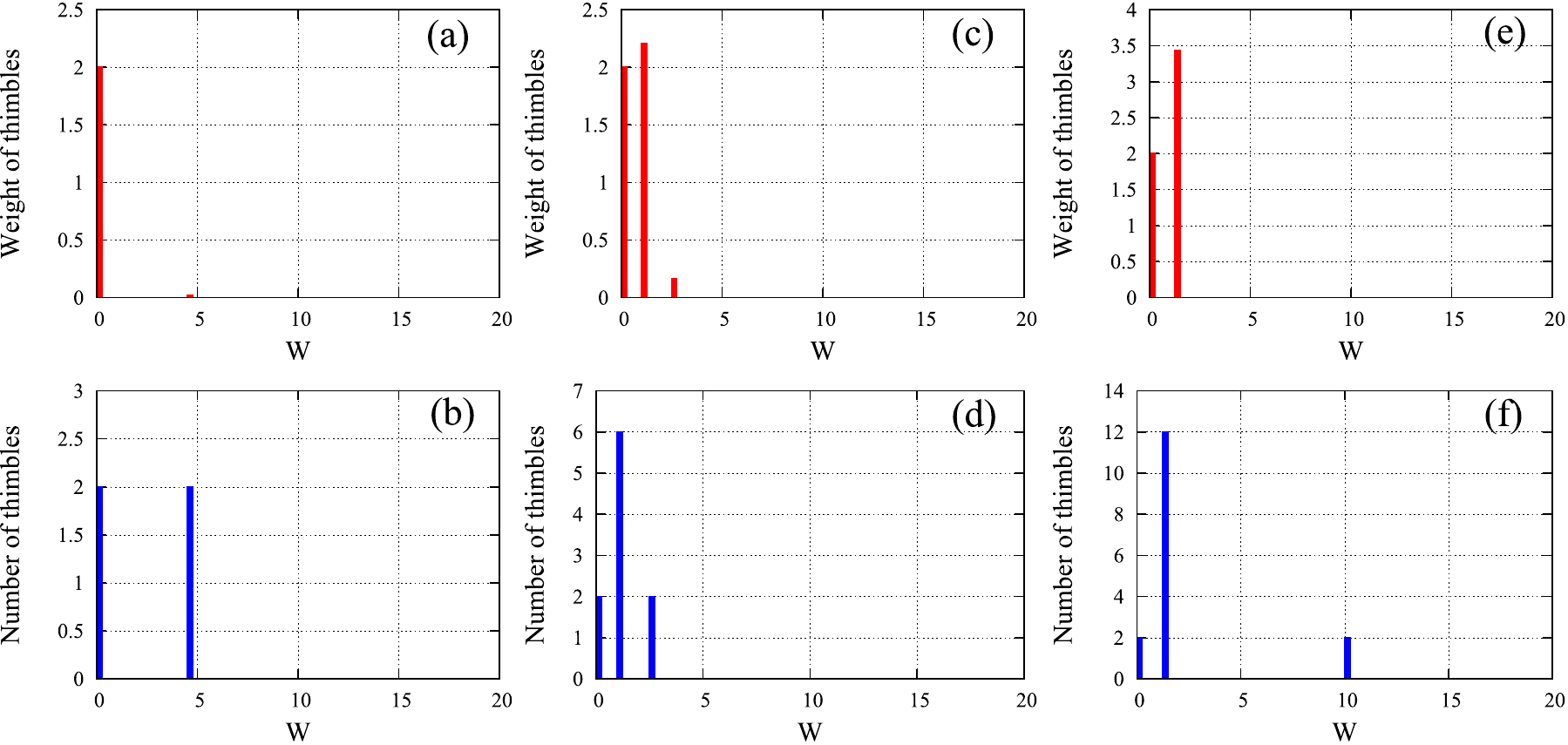}
        \caption{Weighted and normal histograms for relevant thimbles at finite chemical potential $\mu \beta=3$ for two-site lattice with $N_t=1$ (a, b) and $N_t=2$ (c, d), and for four-site lattice with $N_t=1$ (e, f) .  Gaussian representation with purely real exponents is used, the action is written according to equations (\ref{Z_continuous}), (\ref{M_continuous}), (\ref{2site_hamiltonian}) and (\ref{4site_hamiltonian})  with parameters: $U\beta=15.0$, $\kappa\beta=3.0$, $\alpha=0$. Weight of thimbles is shown with respect to the vacuum one.}
        \label{fig:2D_2DNt2_4D_gauss_real_weight_mu3.0}
%saddles_half_filling_2D_Nt1_gauss.nb
\end{figure*}

\subsection{Gaussian HS transformation with only real exponents}

Now we check the same lattices using HS transformation with only real exponents. Action is again constructed according to (\ref{Z_continuous}) and (\ref{M_continuous}) but parameter $\alpha$ is equal to zero, thus only fields $\chi_{x,t}$ remain in the integral. The figure \ref{fig:2D_4D_6D_gauss_real_weight} shows the weight of relevant thimbles at half filling for two- and four-site lattices with $N_t=1$. The situation is much better in comparison with the analogous distributions for the action with complex exponents (fig. \ref{fig:2D_4D_gauss_weight}). The fermionic determinant is no more periodical function of the fields in $\mathbb{R}^N$, thus there is always finite number of relevant thimbles.

\begin{figure}
        \centering
        \includegraphics[scale=0.25]{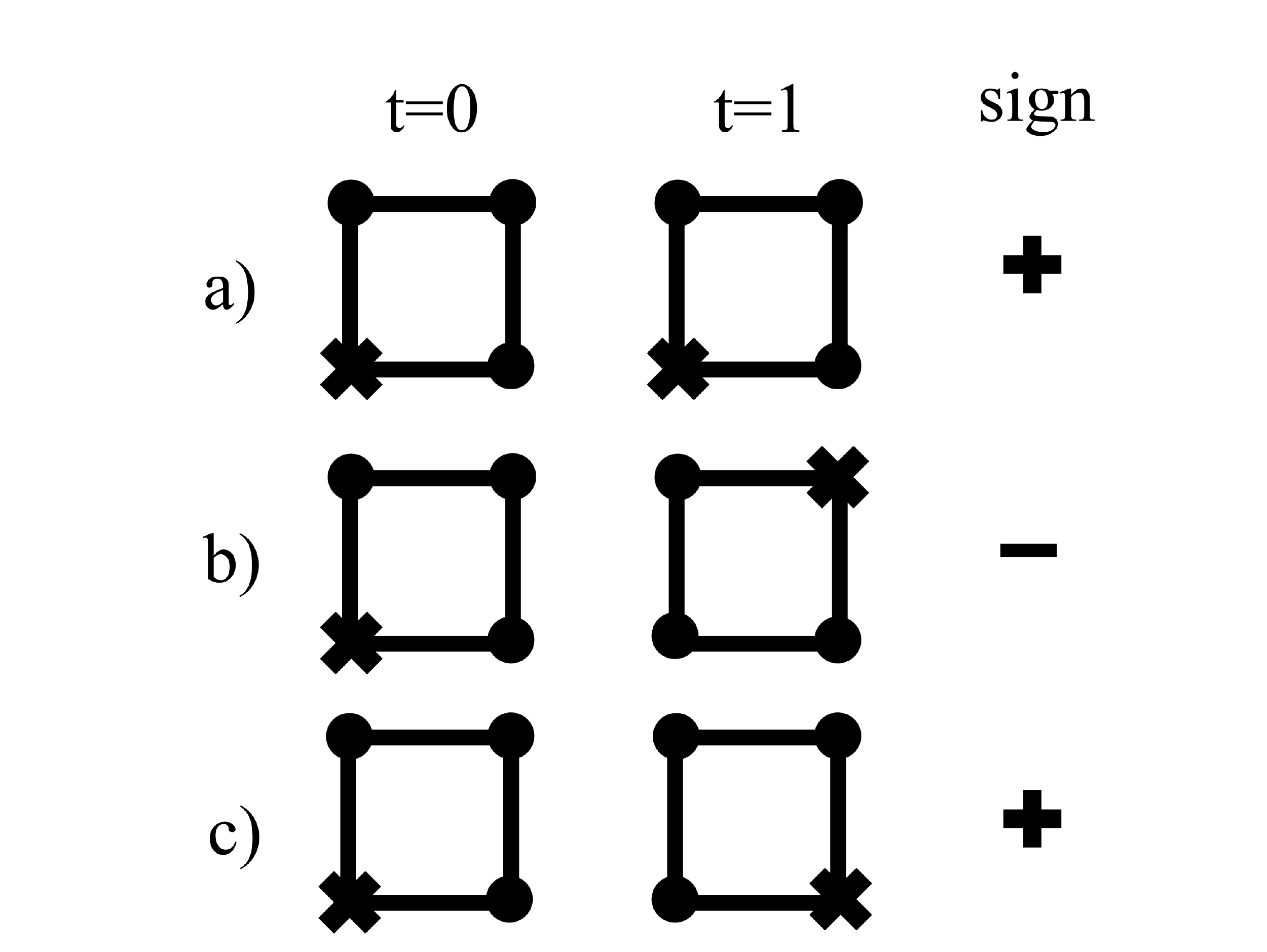}
        \caption{Examples of configurations on the lattice with $N_s=4$, $N_t=2$, with only real exponents in the action. The scheme shows three configurations, two of them have positive weight ((a) and (c)) and one has negative weight (b). Both time slices are shown in each case. Circles correspond to the Hubbard field $\chi \approx - U\delta$ and crosses correspond to $\chi \approx U\delta$, $\delta=\beta/2$. }
        \label{fig:conf_examples}
\end{figure}

However, the situation with the number of dominant thimbles is still rather complicated. The two thimbles with the lowest weight correspond to the saddle points where the Hubbard fields $\chi_{x,t}$ have opposite values at two sublattices but are uniform within each sublattice. The two saddles with the smallest weight (the last bar in the figures \ref{fig:2D_4D_6D_gauss_real_weight}\textcolor{red}{b} and  \ref{fig:2D_4D_6D_gauss_real_weight}\textcolor{red}{d}) correspond to uniform field configurations. And finally, the large number of saddle points with highly non-uniform Hubbard field configurations appear with increased $N_s$ and $N_t$ (see the bar in the middle in the figures \ref{fig:2D_4D_6D_gauss_real_weight} and \ref{fig:2D_Nt2_Nt3_gauss_real_weight}). These non-uniform saddles dominate in the sum over thimbles for all lattices studied in our paper except of the smallest one with $N_s=2$ and $N_t=1$. More detailed analysis of the continuous limit is presented in the figure \ref{fig:2D_Nt2_Nt3_gauss_real_weight} for the lattice with $N_s=2$ and $N_t=2,3,4$. Different types of saddle points are described symbolically in the inset in the figure \ref{fig:2D_4D_6D_gauss_real_weight}\textcolor{red}{f}: we always have two lowest saddles with Hubbard fields uniform within sublattices and two highest saddles with Hubbard fields uniform in entire lattice. In between of them we have a set of saddles where Hubbard fields fluctuate between sublattices. Their weight is very close to each other, only small splitting appears for $N_t=4$. Thus their overall number can be estimated as $2^{N_t}-2$, taking into account all possible reflections and translations. The competition arises between exponentially increased amount of these thimbles and their exponentially decreasing weight (this fact can be also noticed from the figure \ref{fig:2D_4D_6D_gauss_real_weight}). It means that in continuous limit $N_t \rightarrow \infty$ these non-uniform saddles can still make significant contribution in the sum (\ref{thimbles_sum}).

\begin{figure} 
        \centering
        \includegraphics[scale=0.5, angle=0]{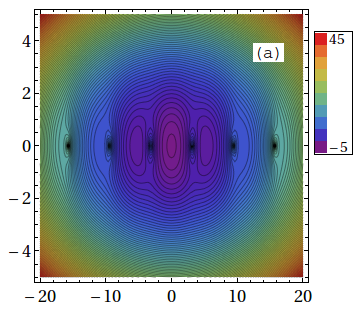}
        \includegraphics[scale=0.5, angle=0]{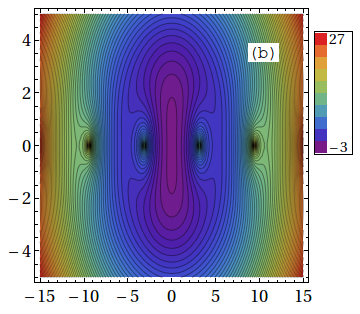}
        \includegraphics[scale=0.5, angle=0]{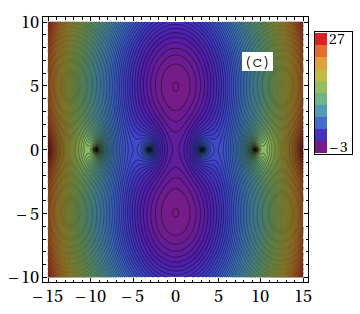}
        \caption{ The action for the one-site Hubbard model in the Gaussian representation: (a) $\alpha=0.95$,   (b) $\alpha=0.8$,  (c) $\alpha=0.5$. $U\beta=10.0$, the action is written according to (\ref{Z_continuous}), (\ref{action_continuous}). Horizontal axis corresponds to the field $\phi$ and vertical axis corresponds to the field $\chi$. Relevant saddle points looks like local minima on the current plots which show only the behaviour of action within $\mathbb{R}^N$. All ``negative'' directions for those saddle points are pointing out in complex space.
       }
        \label{fig:action_combined}
\end{figure}

The case with non-zero chemical potential is shown in the figure \ref{fig:2D_2DNt2_4D_gauss_real_weight_mu3.0}. Unlike the previous study for complex exponents, where we could not directly find relevant complex saddles thus we just tracked the evolution of former real saddles, here we can find all relevant saddles exactly, because all of them are lying within $\mathbb{R}^N$ even for $\mu \neq 0$. The figure \ref{fig:2D_2DNt2_4D_gauss_real_weight_mu3.0} shows the situation for $\mu \beta =3$. It shows three different cases: 1) $N_s=2$, $N_t=1$; 2) $N_s=2$, $N_t=2$; 3) $N_s=4$, $N_t=1$. Thus we can trace both trends: increasing $N_t$ and increasing $N_s$. Comparison with the figures \ref{fig:2D_4D_6D_gauss_real_weight} and \ref{fig:2D_Nt2_Nt3_gauss_real_weight} shows that there are only small changes in the distribution of relevant thimbles.

Importantly, all relevant saddles still have positive weight. This is not surprising taking into account that the absence of the sign problem for the two-site model with Hamiltonian (\ref{2site_hamiltonian}) can be proven analytically even in the continuous limit $N_t\rightarrow \infty$. It means that the sign problem for the path integral representation with purely real exponents appears only if we increase $N_t$ and $N_s$ further. In order to give a short overview how the sign problem looks like in this case, we studied the lattice with $N_s=4$ and $N_t=2$. Indeed, the saddle points with negative sign appear in this case. There is, of course, very large amount of various non-uniform saddles on this lattice, so we just give some examples. The two saddle points with auxiliary fields being uniform across sublattices are again the lowest ones and they still have positive weight. However, the non-uniform saddles with smaller weight can change the sign. Typically, those saddle points which are non-uniform only in space but uniform in the Euclidean time direction have positive weight. On the other hand, the saddle points, which are non-uniform both in space and time, can have both positive and negative weight and the cancellation appears between them. Example configurations for the saddle points with positive and negative weight are shown in the figure \ref{fig:conf_examples}. The observation that the field configurations, which are highly non-uniform in Euclidean time, are responsible for the sign problem is in line with early observations made for BSS-QMC \cite{BSS}. 

In general, the sign problem in the limit $\alpha=0$ is much milder in comparison with the $\alpha=1$ case. This is true at least for small lattice studied so far. The reason is the finite number of relevant thimbles for $\alpha=0$ and much smaller fluctuations of the phase factors for these thimbles.

\subsection{Gaussian HS transformation in mixed regime}

Now we explore the ``mixed'' regime where $\alpha \in (0,1)$ and both real and complex exponents appear in the action (\ref{action_continuous}). At the moment we know that the sign problem is milder in the $\alpha=0$ case, but we will explicitly demonstrate below that the number of relevant thimbles can be reduced even further by tuning the parameter $\alpha$. Additional argument in favor of the mixed regime is the appearance of ``domain walls'' in both limits. Thus neither $\alpha=1$ nor $\alpha=0$ is entirely suitable for the simulations even at half filling. 

We start from the simplest example of the one-site Hubbard model. Figure  \ref{fig:action_combined} shows the action (including logarithms of fermionic determinants)  constructed according to eq. (\ref{Z_continuous}),(\ref{M_continuous}),(\ref{action_continuous}) for one-site model. It is plotted as a function of two fields $\phi$ and $\chi$ for different values of $\alpha$.     
 In the limit when $\alpha$ approaches 1 we have infinite number of relevant saddle points located along the $\phi$-axis and separated by the barriers around the points where the determinant is equal to zero. The vacuum thimble is dominant but the others still have significant weight. Once $\alpha$ decreases,
  the non-vacuum saddles emerged due to the periodicity of the fermionic determinant as a function of $\phi$ start to disappear. If $\alpha < 0.86$ we arrive at the intermediate regime where only vacuum saddle point is relevant. The two saddles along the $\chi$ axis appear only if  $\alpha<0.84$. If $\alpha$ decreases further, these saddle points become more and more disconnected. Thus we have the situation when the length of the HMC trajectory should become longer and longer in order to ensure the possibility to visit both saddle points which are equally important in the probability distribution. This situation is again very disadvantageous for QMC calculations. 
  
Summarizing, in the interval $\alpha \in [0.84, 0.86]$ we have only one relevant thimble at half filling for the one-site Hubbard model. Even more generally, the vicinity of the vacuum configuration $\phi=\chi=0$ is strongly dominant if $\alpha$ is around this interval because the saddles along the $\phi$-axis are lifted or even disappeared and the two saddles along $\chi$- axis (if appeared) are still very close to the vacuum.  Important point is that the ``sweet spot'' regime exists also for larger lattices. For example, at the lattice with $N_s=2$, $N_t=1$ all saddles connected with $\phi$-periodicity of the fermionic determinant disappear at $\alpha<0.92$. On the other hand, the sub-dominant thimbles corresponding to the second bar in the histogram for the case with real exponents (see fig. \ref{fig:2D_4D_6D_gauss_real_weight}\textcolor{red}{a} and \ref{fig:2D_4D_6D_gauss_real_weight}\textcolor{red}{b}) appear only if $\alpha<0.63$. In between of these two values of $\alpha$ we have the regime with only two relevant saddle points at half filling. These saddles are marked in the figure \ref{fig:2D_4D_6D_gauss_real_weight}\textcolor{red}{b}. The situation repeats also at the lattice with $N_s=4$, $N_t=1$. Due do increased system size, we didn't scan the whole interval  $\alpha \in [0,1]$, but at least at $\alpha=0.9$ there are again only two relevant saddles (the first bar in the figure \ref{fig:2D_4D_6D_gauss_real_weight}\textcolor{red}{d}) which correspond to auxiliary fields $\chi$ being uniform across sublattices, with different signs at different sublattices. Auxiliary fields $\phi$ are zero at these saddles. 
In the next section we present also some numerical observations that the regime with reduced number of relevant thimbles exists also in more realistic situations with large $N_t$.

\begin{figure}
        \centering
        \includegraphics[scale=0.6]{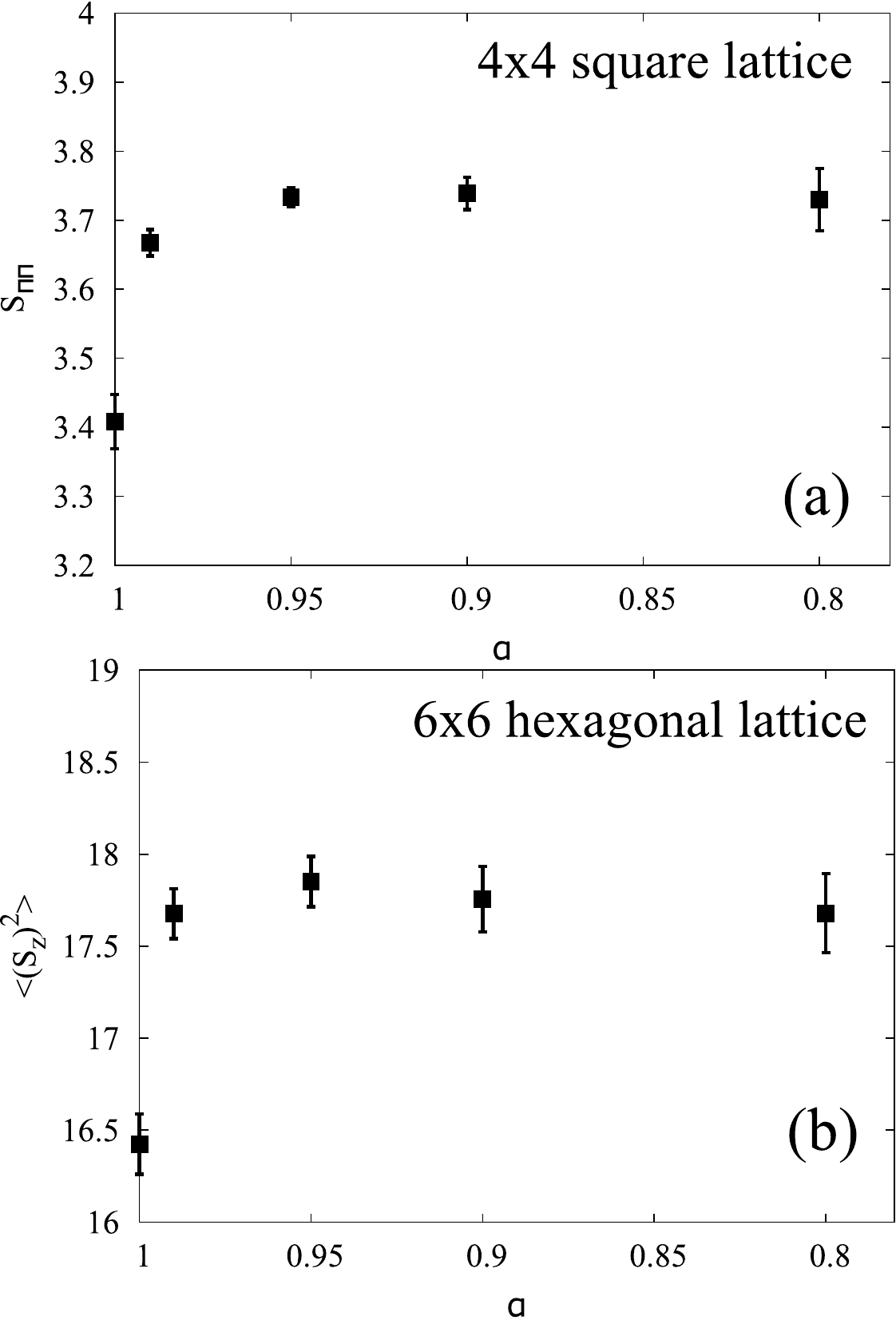}
        \caption{(a) The dependence of the spin structure factor $S_{\pi \pi}$ on $\alpha$ for the $4\times4$ square lattice. The following parameters were used for the simulation: $U=4\kappa$, $N_t=320$, $\beta\kappa=40$. 
        (b) The dependence of the squared spin at one sublattice on $\alpha$ for the $6\times6$ hexagonal lattice. Parameters of the simulation: $U=3.8\kappa$, $N_t=128$, $\beta\kappa=20$.  }
        \label{fig:spin_alpha}
\end{figure}

\section{\label{sec:real_simulations}Results from test HMC simulations}

In order to expand the results from the previous section to more practical cases, we performed some simulations using HMC at different values of $\alpha$. Two settings were used: 1) square $4\times4$ lattice with $N_t=320$ time slices in Euclidean time, on-site interaction $U=4\kappa$, and inverse temperature $\beta\kappa=40$; 2) $6\times6$ lattice with $N_t=128$ time slices, on-site interaction $U=3.8\kappa$, and inverse temperature $\beta\kappa=20$. 
In the case of the square lattice we calculate the spin structure factor
\begin{equation}
  S_{\pi,\pi} = \frac{4}{N} \sum_{i,j} e^{i \vec Q (\vec R_i-  \vec R_j) }   \langle \hat S^i_z   \hat S^j_z  \rangle,  
  \label{spin_formfactor}
\end{equation}
where $\vec Q=(\pi, \pi)$, $\vec R_i$ is the coordinate of the i-th site of the lattice. The squared spin per sublattice is calculated for the hexagonal lattice:
\begin{equation}
  \langle (S_z)^2 \rangle  = \langle \left({ \sum_{i \in 1st. sublat.}  \hat S^i_z  }\right)^2  \rangle.
  \label{spin_squared}
\end{equation}
Results are presented in the figure \ref{fig:spin_alpha}. The effect of the elimination of ``domain walls'' existing in the limit $\alpha=1$ is clearly seen, since the calculations made at $\alpha=0$ show noticeably different results. However, the overall role of the thimbles behind the ``domain walls'' is rather small and doesn't exceed 10 \%. The comparison with the figure \ref{fig:2D_4D_gauss_weight} indeed supports the previous claim that the weight of all additional saddles appearing due to the periodicity of the fermionic determinant as a function of fields $\phi$ decreases in continuous limit. Even quite small value of $\alpha \sim 0.05$ is already enough to eliminate the problems with ergodicity.    

\begin{figure}
        \centering
        \includegraphics[scale=0.6]{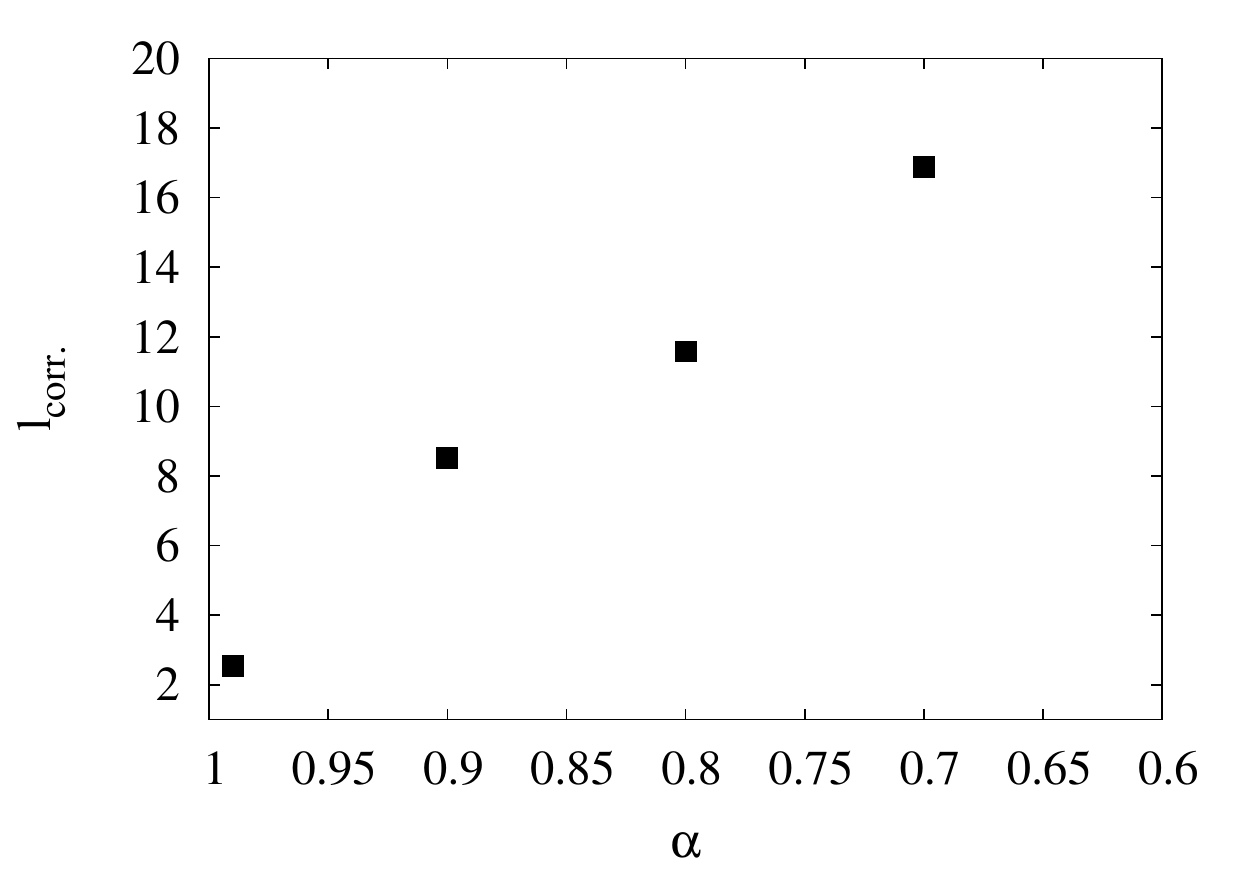}
        \caption{ Correlation length characterizing the autocorrelation between configurations in the HMC process. The runs for different $\alpha$ are equivalent in a sense that the trajectory length in Hamiltonian updates is fixed.  One can see that the autocorrelation grows signalling about the problems with ergodicity for smaller values of $\alpha$. Results are shown for $6\times6$ hexagonal lattice, with the following parameters: $U=3.8\kappa$, $N_t=128$, $\beta\kappa=20$.
        }
        \label{fig:corr_length}
\end{figure}

Figure \ref{fig:corr_length} shows the dependence of the autocorrelation between configurations of auxiliary fields on the parameter $\alpha$. Since all parameters of HMC process including the trajectory length are fixed, increased correlation length means that configurations tend to stuck within some regions of the phase space when $\alpha$ decreases. It directly corresponds to the phenomena we observed in the previous section for one-site and few-site models (see e g. figure \ref{fig:action_combined}\textcolor{red}{c}), where the relevant saddle points tend to form more and more separated peaks in the probability distribution with decreasing $\alpha$. Combining this observation with the results for observables (figure \ref{fig:spin_alpha}) we can conclude that the ``sweet spot'' regime exists also in these more realistic calculations at approximately the same values of $\alpha \approx 0.9$. In this regime we already get rid of the ``domain walls'' but the probability distribution for the fields $\phi$ and $\chi$ is not yet separated into distinct peaks.  

\begin{figure}
        \centering
        \includegraphics[scale=0.6]{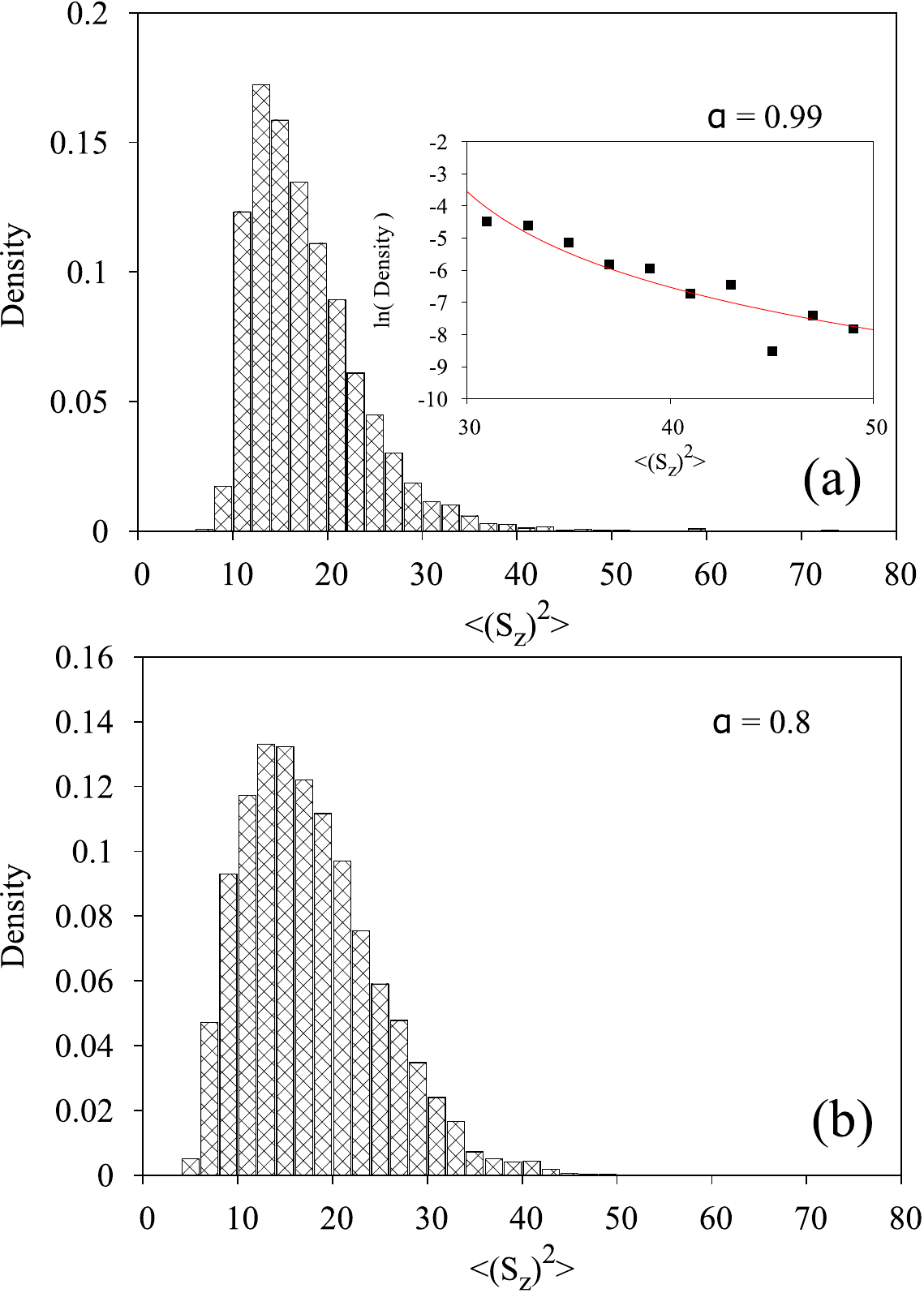}
        \caption{ Histograms showing the distribution of the observable (squared  spin on sublattice) in different  Monte Carlo field configurations. Results are presented for $6\times6$ hexagonal lattice, with the following parameters: $U=3.8\kappa$, $N_t=128$, $\beta\kappa=20$, $\alpha=0.99$ and 0.8. The inset in the figure (a) 
         shows the logarithm of the distribution  fitted with the function $\ln{A/(x-B)^{2.5}}$.
        }
        \label{fig:spin_hist}
\end{figure}

Unfortunately, the existence of the field configurations with zero fermionic determinant still introduces important obstacles in QMC algorithms for continuous auxiliary fields even if  the problems with ergodicity are absent. Namely, the distributions of  observables become heavy tailed with indefinite second moment. Thus the standard statistical post-processing based on the central limit theorem may not be applicable. Let us consider the complementary cumulative distribution function for the spin structure factors $\Sigma$. Among other terms, they always include squared Euclidean fermionic propagator $g^2$. This is the most important term in the vicinity of the configurations where the fermionic determinant is zero, since the propagator diverges there as $1/\Delta$.  $\Delta$ is the distance to the zero point of the determinant in the space of auxiliary fields $\phi$ and $\chi$. 
Due the divergence of observables near the zero points of the determinant, exactly these field configurations define the complementary cumulative distribution function  $\bar F_\Sigma (S)$ at large values of observable. If $\alpha=1$ (only $\phi$ fields appear in the integrals) the asymptotic behaviour of the function $\bar F$ can be described by the integral: 
\begin{equation}
    \bar F_{\Sigma}(S)|_{S \rightarrow \infty} =\mathcal{P} (\Sigma>S) =\int_{V:\Sigma>S} d^N \phi P (\phi),
    \label{cumul_distr1}
\end{equation}
where $P(\phi)$ is the probability distribution for the $\phi$ fields. If $S$ is sufficiently large, the volume $V$ is just some thin layer in the vicinity of the ``domain wall'', where  $P(\phi)=0$.  Now we change the variables so that $x_2...x_N$ correspond to the shift parallel to the ``domain wall'' while the coordinate $x_1$ is perpendicular to it. The ``domain wall'' itself corresponds to $x_1=0$. Thus
\begin{equation}
    \bar F_{\Sigma}(S)|_{S \rightarrow \infty} =\int_{V:\Sigma>S} d^N x \frac{\mathcal{D}(\phi)}{\mathcal{D}(x)} x_1^2 f(x_2...x_N),
     \label{cumul_distr2}
\end{equation}
and we used eq. (\ref{M_square_cont}) in order to estimate the probability distribution $P(x)$ in the vicinity of ``domain walls'' as $P(x) \approx x_1^2 f(x_2...x_N)$. Since the observable diverges as we approach the ``domain wall'':
\begin{equation}
    \Sigma\approx \frac{C(x_2, ... x_N)}{x_1^2},
\end{equation}
the integral over $x_1$ in (\ref{cumul_distr2}) has the limits
\begin{equation}
x_1 \in [-C_1(x_2, ... x_N)/\sqrt{S},C_2(x_2, ... x_N)/\sqrt{S} ],  
\end{equation}
where $C_1,C_2>0$. If the Jacobian doesn't have any divergences in the limit $x_1\rightarrow 0$, the integral over $x_1$ in (\ref{cumul_distr2}) can be computed separately. Thus the asymptotic behaviour of the function $\bar F$ is described by the expression:
\begin{eqnarray}
    \bar F_{\Sigma}(S)|_{S \rightarrow \infty} \approx \frac{\mathcal{C} }{S^{3/2}}, \\
     \mathcal{C} =  \frac{1}{3}  \int dx_2 ... dx_N \left.{\frac{\mathcal{D}(\phi)}{\mathcal{D}(x)}}\right|_{{x_1}=0}\times \\ \times \left({C_1(x_2,...x_N)}^3 + {C_2(x_2,...x_N)}^3\right). \nonumber
     \label{cumul_distr_fin1}
\end{eqnarray}
Conversion to the probability distribution gives
\begin{eqnarray}
    P_{\Sigma}(S)|_{S\rightarrow \infty} \sim \frac{1}{S^{5/2}}.
    \label{asympt1}
\end{eqnarray}
The same derivation can be repeated in the opposite limit of purely real exponents ($\alpha=0$). 
Away of these two limits the derivation has to be modified since the dimensionality of the manifolds with zero determinant is reduced to $N-2$, where $N$ is the total number of auxiliary fields $\phi$ and $\chi$.  We should now separate two coordinates $x_1$ and $x_2$ which corresponds to the shift perpendicular to the ``line'' with zero fermionic determinant, while all other coordinates $x_3,..x_N$ again correspond to the shift parallel to this ``line''. After it $x_1$ and $x_2$ are changed to polar coordinates $(\rho, \varphi)$ and the resulting asymptotic behaviour for the complementary cumulative distribution function is described by the expression
\begin{eqnarray}
    \bar F_{\Sigma}(S)|_{S \rightarrow \infty}  \approx  \int d\varphi dx_3 ... dx_N \\ \int_0^{C(\varphi,x_3,..x_N)/\sqrt{S}} \rho^3 d\rho   \left.{\frac{\mathcal{D}(\phi)}{\mathcal{D}(x)}}\right|_{{\rho}=0} f(\varphi, x_3...x_N). \nonumber
     \label{cumul_distr_fin1}
\end{eqnarray}
Additional power of $\rho$ appears from the transfer to polar coordinates. In principle, this power appears from the Jacobian if the transfer to polar coordinate is included in the general change of variables. The probability distribution for the spin structure factors has now the asymptote
\begin{eqnarray}
    P_{\Sigma}(S)|_{S\rightarrow \infty} \sim \frac{1}{S^3}.
    \label{asympt2}
\end{eqnarray}
Similar derivations can be repeated for lower dimensions of the manifolds with zero fermionic determinant with larger powers of $\rho$ appearing from the Jacobian. Generally, the lower the dimensionality of the manifolds with zero fermionic determinant leads to the larger absolute value of the power in the tail of the distribution. 

In order to demonstrate this effect explicitly, we plotted the distributions of the observable (squared spin on sublattice) obtained from the calculations on hexagonal lattice.
Results are shown in the figure \ref{fig:spin_hist}. The heavy tailed distribution is clearly seen both for $\alpha=0.99$ and $\alpha=0.8$. The inset in the figure \ref{fig:spin_hist}\textcolor{red}{a} shows that the tail of the distribution can not be described by the Gaussian function. The quality of data is not good enough to define the power law from the fitting, but the logarithm of the distribution definitely can not be fitted with $-(x-x_0)^2/2\sigma$ with any reasonable dispersion $\sigma$. 

One can see that in both cases (\ref{asympt1}) and (\ref{asympt2}) the second moment for the observable is not defined. It means that the standard estimation of error based on the calculations of dispersion is not applicable. Correct procedure must include data binning with numerical estimation for the confidence interval on the basis of the final distribution for the averages.

\section{\label{sec:new_1site}Non-Gaussian representation for the interaction term}

\begin{figure*}
        \centering
        \includegraphics[scale=0.7]{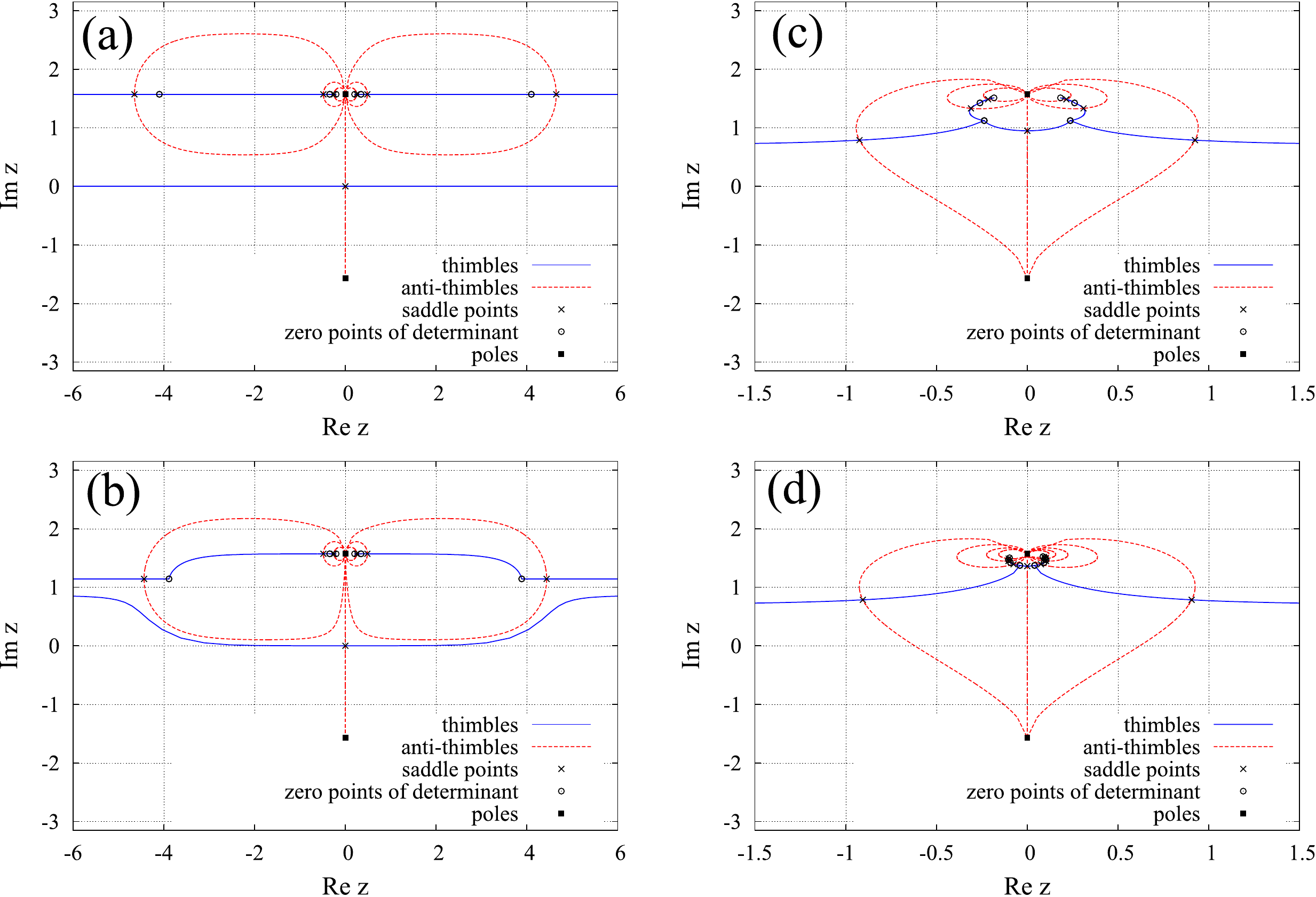}
        \caption{Thimbles and anti-thimbles for the one-site Hubbard model in the non-Gaussian representation at various values of chemical potential. The action is written in (\ref{Z_new_final}). $U\beta=15.0$. (a)  Half filling ($\mu=0$).  Only one thimble covers the whole real axis. (b) Small chemical potential ($\beta \mu=0.002$). The Stokes phenomenon hasn't yet occurred and we have still only one relevant saddle: the former ``vacuum'' one but shifted to complex plane. (c) Intermediate values of chemical potential ($\beta \mu=5.0$). The Stokes phenomenon has already occurred and we have three relevant saddles. However, the statistical weight of former ``vacuum'' saddle still prevails. (d) Large chemical potential ($\beta \mu=15.0$). There are still three relevant thimbles, but the statistical weight of former ``vacuum'' saddle is negligible in comparison with two symmetrical complex saddles. In all cases anti-thimbles corresponding to irrelevant saddles have both their ends in the same pole $\mbox{Im}\, z=\pi/2, \, \mbox{Re}\,z=0$. We show only several saddles close to the point $\mbox{Im}\, z=\pi/2, \, \mbox{Re}\,z=0$, in fact there is infinite number of saddles approaching the pole.}
        \label{fig:one_site_new_thimbles}
%thimbles_gauss_full.nb
\end{figure*}

 \subsection{One-site Hubbard model}

In this section we derive the non-Gaussian integral representation of the exponents with four-fermionic terms. We choose the integral representation because it preserves the Cauchy's Integral theorem and the possibility to reduce the sign problem by shifting the integration contour to the complex plane. The main motivation is to demonstrate an alternative way to get rid of the infinite number of relevant thimbles appearing in the Gaussian HS transformation with only complex exponents. 

We start from the integral representation with overall structure similar to BSS-QMC (\ref{discrete_HS}) with the auxiliary fields bounded to some finite region. The interaction term  in  (\ref{Hamiltonian_el_hol}) can be written as the series:
\begin{eqnarray}
    \label{exp_series}
e^{-\frac{\delta U}{2} (\hat n_{el.} - \hat n_{h.})^2 }  = \nonumber \\ = 1+ \sum_{j=1}^{\infty} \frac{(-\delta U /2)^j}{j!} (\hat n_{el.} - \hat n_{h.})^2
\end{eqnarray}
On the other hand
\begin{eqnarray}
    \label{new_trans_series}
\frac{1}{2} \int_{-1}^{1} e^{i \gamma x (\hat n_{el.} - \hat n_{h.})} dx = \nonumber \\ 
= 1+  (\hat n_{el.} - \hat n_{h.})^2 \int_0^1 \left({ \sum_{m=0}^\infty \frac{(-1)^m (\gamma x)^{2m}}{(2m)!} } \right) dx.
\end{eqnarray}
Thus the following integral transformation is possible:
\begin{equation}
    \label{new_HS}
  e^{-\frac{\delta U}{2} (\hat n_{el.} - \hat n_{h.})^2 } = \frac{1}{2} \int_{-1}^{1} e^{i \gamma x (\hat n_{el.} - \hat n_{h.})} dx,  
\end{equation}
where constant $\gamma$ is defined as
\begin{equation}
    \label{new_HS_gamma}
e^{-\frac{\delta U}{2}} = \frac{\sin \gamma}{\gamma}.
\end{equation}
In some cases this equation has several physically equivalent solutions. In this paper, we always get the smallest possible value of $\gamma$. Using this formula, we arrive to the path integral representation of the partition function very similar to (\ref{Z_discrete}) and (\ref{M_discrete}):
\begin{eqnarray}
  \mathcal{Z}_n = \int_{-1}^1 \mathcal{D}\psi_{x,t} \det P_{el.}(\psi_{x,t}) \det P_{h.}(\psi_{x,t})
  \label{Z_new}
\end{eqnarray}
with fermionic operators for electron and holes defined as:
\begin{eqnarray}
 P_{el.}(\psi_{x,t}) = I +\prod^{N_t}_{t=1} \left({ e^{-\delta (h+\mu)} \diag{ e^{i \gamma \psi_{x,t}} } }\right), \nonumber \\
 P_{h.}(\psi_{x,t}) = I +\prod^{N_t}_{t=1} \left({ e^{-\delta (h-\mu)} \diag{ e^{-i \gamma \psi_{x,t}} } }\right). 
  \label{M_new}
\end{eqnarray}
As in the previous cases $\left({ \diag{ e^{i \gamma \psi_{x,t}} }}\right)$ is a $N_s \times N_s$ diagonal matrix, which contains the exponents of auxiliary fields belonging to one time slice.

In principle, one can work directly with this representation for the partition function changing the fields $\psi_{x,t}$ locally or organizing the Hamiltonian updates in the bounded region. 
However, the organization of dynamics for the field bounded within the interval $\psi_{x,t} \in [-1;\,1]$  needs some modifications of the HMC algorithm. The special ``reflection'' steps should be made at the border. This can be avoided if we change the variables stretching the integration domain back to infinity:
\begin{equation}
\label{change_var}
\psi_{x,t} = \tanh z_{x,t}, \quad z_{x,t}\in(-\infty, \, +\infty).
\end{equation}
Effectively we introduce ``soft walls'' on the borders of initial integration domain $[-1;\,1]$. It also allows us to treat the integration contours in a similar manner as in the case of Gaussian HS transformation. For example, we do not need to attach the shifted contours to the points $\pm 1$ at real axis in order to preserve the value of the integral. 
Here is the final form of the partition function:
\begin{eqnarray}
  \mathcal{Z}_n = \int_{-\infty}^{+\infty} \mathcal{D} z_{x,t} e^{-S_{n} (z_{x,t})}, \nonumber \\
  S_{n} (z_{x,t}) = 2 \sum_{x,t} \ln \cosh z_{x,t} - \nonumber \\
  - \ln \left( { \det P_{el.}(\tanh z_{x,t}) \det P_{h.}(\tanh z_{x,t})}\right) 
  \label{Z_new_final}
\end{eqnarray}
The introduction of the hyperbolic functions does not make the calculation much more expensive then the Gaussian approach (\ref{general_action}), because we need to compute the complex exponents in fermionic determinants (\ref{M_continuous}) and (\ref{M_new}) in any case. Thus, additional calculation of the exponent $e^z$ needed for the evaluation of hyperbolic functions doesn't introduce significant additional difficulty in the algorithm. However, we should mention that this change of variables might be not the most efficient choice. Probably, some other variants should be tested. 

This representation suffers from the same ergodicity problem (see eq. (\ref{M_zero_modes_relation})) which appears if we use Gaussian HS transformation with only complex exponents: domain walls formed by configurations with zero fermionic determinant. It can be solved using the same trick with combination of real and complex exponents. The same derivation for real exponents gives us
\begin{equation}
    \label{new_HS_real}
  e^{-\frac{\delta U}{2} (\hat n_{el.} - \hat n_{h.})^2 } = \frac{1}{2} \int_{-1}^{1} e^{\lambda x (\hat n_{el.} - \hat n_{h.})} dx,  
\end{equation}
where constant $\lambda$ is defined as
\begin{equation}
    \label{new_HS_lambda}
e^{\frac{\delta U}{2}} = \frac{\sinh \lambda}{\lambda}.
\end{equation}
So we can combine it with (\ref{new_HS}) and get rid of the ``domain walls''.

Now we study the one-site Hubbard model using the new path integral representation with only complex exponents and compare our results with \cite{Tanizaki}. 
According to (\ref{Z_new_final}), the action should be written as
\begin{eqnarray}
 S^{(1)}_{n} (z) = 2 \ln \cosh z - \nonumber \\
  - \ln \left( { (1+e^{i\gamma \tanh z -\beta \mu } ) (1+e^{-i\gamma \tanh z +\beta \mu } )   }\right).
  \label{S_one_site_new}
\end{eqnarray}
Complex saddle points as well as corresponding thimbles and anti-thimbles are shown in the figure \ref{fig:one_site_new_thimbles} for the same interaction strength $U\beta=15.0$ and four different values of the chemical potential. Since the action is periodic along the imaginary axis with period $2 \pi$ (and this property is preserved in the case of full many-site Hubbard model), we can look only at the strip $\mbox{Im}\, z \in [-\pi, \, \pi]$. Unlike the Gaussian case (fig. \ref{fig:one_site_gauss_thimbles}\textcolor{red}{a}), we have only one relevant thimble at half filling which coincides with the whole real axis. An important property of the representation (\ref{Z_new_final}) is that the action has now an additional 
singular points, namely, the poles at $\mbox{Im}\, z= \pm \pi/2, \, \mbox{Re}\,\, z=0$. All anti-thimbles now should end up at one or both of these points. Obviously, anti-thimbles corresponding to relevant saddle points start at one pole and end up at another one with an opposite sign of $\mbox{Im}\, z$.

\begin{figure*}
        \centering
        \includegraphics[scale=0.7]{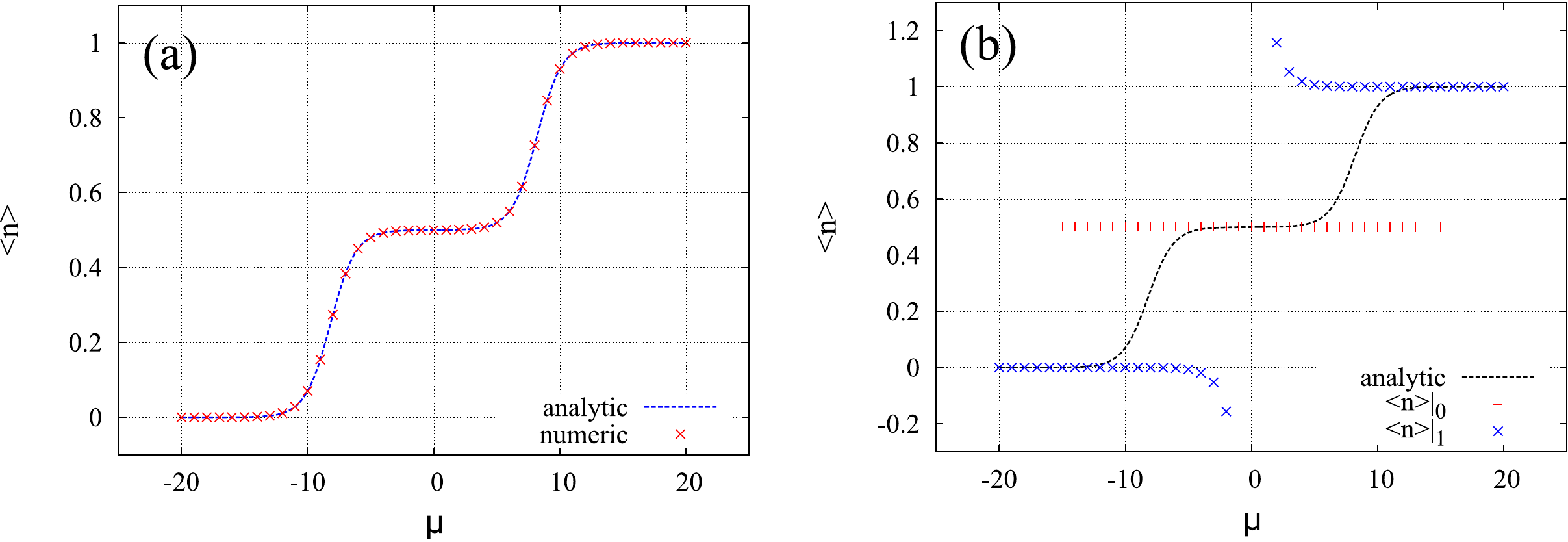}
        \caption{(a) Average number of particles $\langle \hat n \rangle$ computed with the non-Gaussian integral representation for the one-site Hubbard model (\ref{thimbles_n}). (b) Contribution of different thimbles to the average number of particles $\langle \hat n \rangle|_0$ and $\langle \hat n \rangle|_1$ computed with the non-Gaussian integral representation for the one-site Hubbard model (\ref{thimbles_n_sep}). Analytic result is also shown for comparison, $U\beta=15.0$.}
        \label{fig:n_new_all}
%thimbles_Exp_full.nb
\end{figure*}

This behaviour is indeed clearly seen in the figure \ref{fig:one_site_new_thimbles}. Initially, at small chemical potential, we still have  the situation where only the ``vacuum'' saddle point at $z=0$ is relevant. However, at larger values of $\mu$ the Stokes phenomenon happens where anti-thimbles of two initially irrelevant saddle points collide with the former ``vacuum'' saddle point. Subsequently, three anti-thimbles pass from one pole to another crossing the real axis. After the Stokes phenomenon the situation with only three relevant saddles persists even at very large values of chemical potential $\mu>U$. 
This is already quite different situation from the Gaussian case with complex exponents, where the series over thimbles (\ref{thimbles_sum}) is infinite. 
In this sense we are already closer to the Gaussian HS transformation with only real exponents. 

\begin{figure*}
        \centering
        \includegraphics[scale=1.1, trim=-0cm 0 0 0]{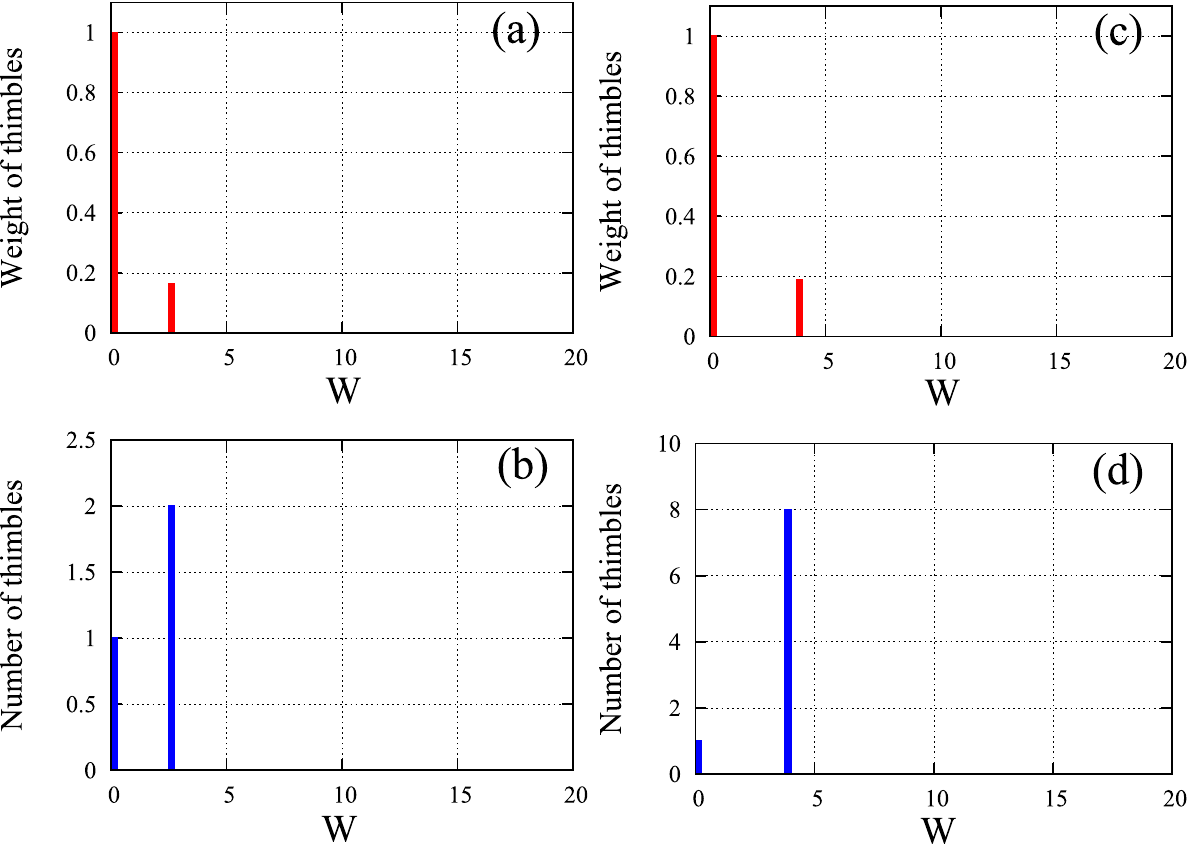}
        \caption{Weighted and normal histograms for relevant thimbles at half-filling for two-site ((a) and (b)), four-site ((c) and (d))  lattice. Non-Gaussian representation is used, $N_t=1$ in all cases.  The action is written according to equations (\ref{Z_new_final}), (\ref{M_new}), (\ref{2site_hamiltonian}) with parameters: $U\beta=15.0$, $\kappa\beta=3.0$. Weight of thimbles is shown with respect to the vacuum one.}
      \label{fig:2D_4D_6D_new_weight}
%saddles_half_filling_2D_Nt1_gauss.nb
\end{figure*}

The role of saddle points in the new representation is clarified in the figure \ref{fig:n_new_all}.  This figure illustrates the contribution to the average number of particles $\langle \hat n \rangle$ from different thimbles.
The full observable is equal to 
\begin{equation}
    \label{thimbles_n}
\langle \hat n \rangle = \frac{\sum_{\sigma=0, \pm 1} \int_{\mathcal{I}_\sigma} d z e^{-S^{(1)}_n} \left({1+e^{i \gamma \tanh z - \mu \beta} }\right)^{-1}}{\sum_{\sigma=0, \pm 1} \int_{\mathcal{I}_\sigma} d z e^{-S^{(1)}_n} },
\end{equation}
where $\sigma=0, \pm 1$ denotes all three relevant thimbles.  The full sum is shown in the figure \ref{fig:n_new_all}\textcolor{red}{a} and indeed, the values obtained via integral over thimbles coincides completely with the analytic answer. The role of different thimbles is revealed if we plot the observable calculated only at given thimbles:
\begin{eqnarray}
    \label{thimbles_n_sep}
\langle \hat n \rangle|_0 = \frac{\int_{\mathcal{I}_0} d z e^{-S^{(1)}_n} \left({1+e^{i \gamma \tanh z - \mu \beta} }\right)^{-1}}{\int_{\mathcal{I}_0} d z e^{-S^{(1)}_n} }, \nonumber \\
\langle \hat n \rangle|_1 = \frac{\sum_{\sigma\pm 1} \int_{\mathcal{I}_\sigma} d z e^{-S^{(1)}_n} \left({1+e^{i \gamma \tanh z - \mu \beta} }\right)^{-1}}{\sum_{\sigma\pm 1} \int_{\mathcal{I}_\sigma} d z e^{-S^{(1)}_n} }.
\end{eqnarray}
$\langle \hat n \rangle|_0$ represents the contribution from ``vacuum'' thimble and $\langle \hat n \rangle|_1$ corresponds to the contribution from two others (they are complex conjugate to each other, thus making equal contribution to the real observable). Figure \ref{fig:n_new_all}\textcolor{red}{b} demonstrates these quantities. One can see that the ``vacuum'' thimble gives a good approximation to $\langle \hat n \rangle$ for small chemical potential while two others work well for $\mu \geq U$ after the transition. We should stress that $\langle \hat n \rangle \neq \langle \hat n \rangle|_0 + \langle \hat n \rangle|_1$ because of different normalization in the denominators of eq. (\ref{thimbles_n}) and (\ref{thimbles_n_sep}).

We see that in this model the phase transition reflects itself in the transfer of statistical weight from one saddle point (former ''vacuum``) to a couple of others which are complex conjugate to each other. Thus, everywhere except the transition region, $\mu \approx U$, there is one (or two complex conjugated) thimbles which provide the dominant contribution to observable. This property when just one thimble (or the group of equivalent thimbles) gives a good approximation for particular phase is very useful in real calculations, because it reduces the problems with ergodicity in QMC. However, we don't know, how this situation will change in the thermodynamic limit, this should be a subject of further study. 

Interestingly, we can now introduce the topological invariant to distinguish relevant and irrelevant saddle points. This possibility relies on the fact that anti-thimbles for irrelevant saddle points have both their ends at one pole, while anti-thimbles for relevant saddles should connect both poles. The invariant for the $\sigma$-th saddle point can be written as
\begin{equation}
    \mathcal{T}_\sigma = \frac{1}{i \pi}\int_{\mathcal{K}_\sigma} dz,
    \label{top_invariant}
\end{equation}
where the integral is taken over corresponding anti-thimble. The invariant is equal to 1 for relevant saddle point and 0 for irrelevant. 
The possibility to generalize this formula to larger dimensions also exists but should be a subject of further study. In principle, the existence of such invariant can help in the detection of relevant saddle points, because we do not need to search for the intersection point, which might be very non-trivial task in many dimensions.

\subsection{\label{sec:Nsite}Few-site Hubbard model}

\begin{figure*}
        \centering
        \includegraphics[scale=1.1, trim=-0cm 0 0 0]{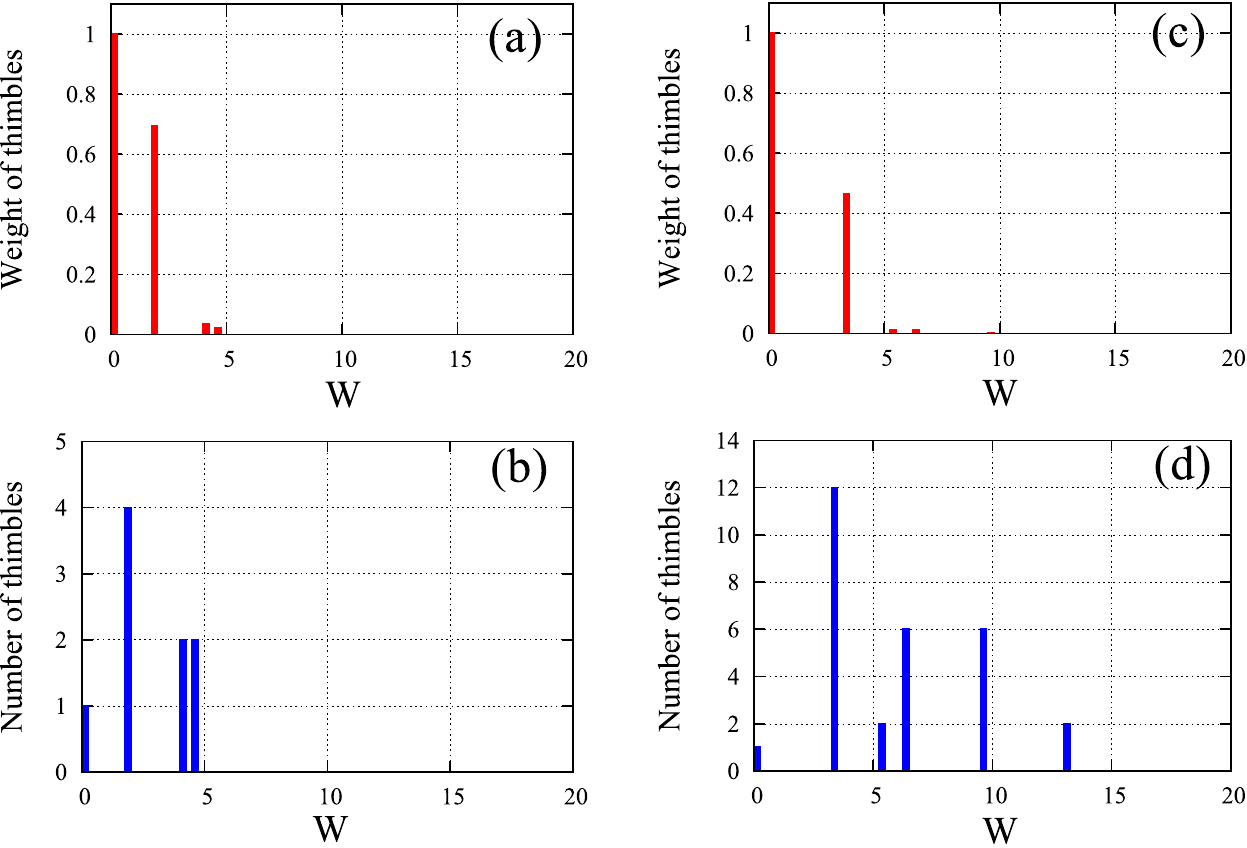}
        \caption{Weighted and normal histograms for relevant thimbles at half-filling for two-site lattice with two ((a) and (b)) and three ((c) and (d)) Euclidean time slices. Non-Gaussian representation is used, the action is written according to equations (\ref{Z_new_final}), (\ref{M_new}), (\ref{2site_hamiltonian}) with parameters: $U\beta=15.0$, $\kappa\beta=3.0$. Weight of thimbles is shown with respect to the vacuum saddle.}
        \label{fig:2D_Nt2_Nt3_new_weight}
%saddles_half_filling_2D_Nt1_gauss.nb
\end{figure*}

Next we study the behaviour of the non-Gaussian integral representation of the interaction term in case of the Hubbard model with few lattice sites. We take the action (\ref{Z_new_final}). In order to make direct comparison, the same parameters ($U\beta=15.0$ and $\kappa\beta=3$) are used as was done for the Gaussian representation. 
The same approach is also used for the estimation of the complexity of the sign problem: we start from half filling and identify all relevant real saddle points; after it we trace their  shift in the complex plane with increasing chemical potential and look at the evolution of their phases.

\begin{figure*}
        \centering
        \includegraphics[scale=0.5]{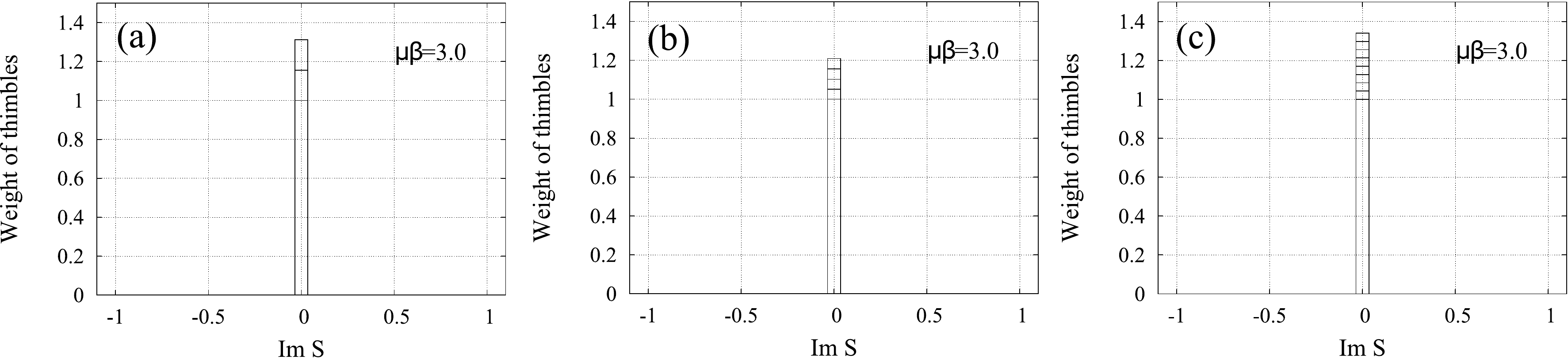}
      \caption{Stacked histograms showing the evolution of real thimbles which were relevant at half-filling. Calculations are made for non-Gaussian representation. Since $\mbox{Im} S$ remains zero for all thimbles, they form only one bar in the histograms. Contributions of these thimbles are separated by horizontal lines within the bar. The weight of former ``vacuum'' thimble is taken as unity.  The calculation is made for two-site lattice with $N_t=1$ (a) and $N_t=2$ (b); and for four-site lattice with $N_t=1$ (c). The action is written according to equations (\ref{Z_new_final}), (\ref{M_new}), (\ref{2site_hamiltonian}) with parameters: $U\beta=15.0$, $\kappa\beta=3.0$, $\kappa\mu=3.0$.  }
        \label{fig:2D_2DNt2_4D_new_weight_mu3.0}
%saddles_new_2D_finite_mu.nb
\end{figure*}

The results for the two- and  four-site lattices  with $N_t=1$ at half filling are shown in the figure \ref{fig:2D_4D_6D_new_weight} in the usual manner (normal and ``weighted'' histogram for relevant real saddles). The most important feature of the new path integral representation is still preserved: instead of an infinite number of relevant saddles we have only finite number of them within the real subspace. Importantly, the ``weighted'' histogram shows that the vacuum saddle still dominates in the sum (\ref{thimbles_sum}). The number of sub-dominant thimbles grows with $N_s$, but their role in the sum (\ref{thimbles_sum}) remains stable, as one can see from the figures  \ref{fig:2D_4D_6D_new_weight}\textcolor{red}{b} and \ref{fig:2D_4D_6D_new_weight}\textcolor{red}{d}. Thus we have similar competition between exponentially increasing number of relevant thimbles and their exponentially decreasing weight as we observed in the case of  the real Gaussian HS transformation. The only difference is that the weight of sub-dominant thimbles is substantially decreased and the ``vacuum'' thimble always dominates. 

The scaling with $N_t$ is shown in the figure \ref{fig:2D_Nt2_Nt3_new_weight} for two-site lattice with $N_t=2$ and $N_t=3$. Comparison of fig. \ref{fig:2D_Nt2_Nt3_new_weight}\textcolor{red}{a} and  \ref{fig:2D_Nt2_Nt3_new_weight}\textcolor{red}{c} shows that that the overall weight of sub-dominant thimbles decreases, despite that their number increases with $N_t$. 

However, these sub-dominant thimbles are not a large issue from the point of view of the sign problem. The results for the former real saddles in the case of increased chemical potential (see fig. \ref{fig:2D_2DNt2_4D_new_weight_mu3.0}) are shown for the following lattices: 1) $N_s=2$, $N_t=1$; 2) $N_s=2$, $N_t=2$; 3) $N_t=4$, $N_t=1$. In all cases we see only saddle points with $\mbox{Im} S=0$. Thus the sign problem is significantly milder in comparison with the case of the Gaussian HS transformation with complex exponents and situation reminds more the Gaussian case with only real exponents.

 \section*{\label{sec:Conclusion}Conclusions}

The Lefschetz thimble analysis of the sign problem was made for the few-site Hubbard model combining the analytic study of the lattices with few Euclidean time slices and the results from HMC simulations on the lattices with large $N_t$, approaching the continuous limit. In this study, we found all relevant saddles in the real subspace $\mathbb{R}^N$ at half filling and looked at their relative importance using the Gaussian approximation to the integrals over thimbles.  After it we track the evolution of these real saddles (including the shift in the complex domain) with increasing chemical potential and look at the phases which they acquire. In principle, some additional relevant saddles can appear or former real saddles can become irrelevant. However, since the action is a continuous function of the chemical potential, additional saddles can not suddenly become dominant. Thus, such an incomplete study still gives a reasonable estimate of the complexity of the sign problem at small values of chemical potential. 

Two variants of the Gaussian HS transformation were studied. They differ by the presence of real or complex exponents with Hubbard fields in the fermionic determinant. Hubbard fields are coupled to particle number operators for electrons and holes. In both cases the complexity of the sign problem (the number of significant thimbles and the fluctuations of their sign) increases with the spatial size of lattice. On the contrary, the continuous limit $N_t\rightarrow \infty$ makes the sign problem milder by lifting the non-uniform saddle points and decreasing their contribution in the overall sum over relevant thimbles. The variant with only real exponents in fermionic determinant exhibits much milder sign problem due to finite number of relevant thimbles and much smaller fluctuations of their phases. 

The minimal number of relevant thimbles is observed when both real and complex exponents are mixed in the action (thus two auxiliary fields per lattice site should be introduced). This regime needs some tuning of the parameter $\alpha$ which defines the relative importance of real and complex exponents in the decomposition of the interaction term. Generally, we should work in the regime closer to the case with only complex exponents ($\alpha \approx 0.9$). In this case the number of relevant thimbles is equal to two for all few-site lattices studied in the paper. Data from HMC test runs also support the claim that this is the most advantageous regime for possible application of Generalized thimble algorithm for the Hubbard model. 

We present also another example of the representation where the number of relevant thimbles is substantially reduced. An alternative integral representation for the interaction term is derived using the analogy with BSS-QMC.  Analysis made on lattices with $N_s=1,2,4$ and $N_t=1,2,3$ shows that the number of relevant thimbles is always finite and the ``vacuum'' saddle point located at zero auxiliary fields is always dominant (at least for moderate values of the chemical potential $\mu \approx  \kappa)$. Moreover, the calculation with finite chemical potential shows that all former real saddles move into complex domain but still preserve zero $\mbox{Im} S$. It means that at small values of chemical potential the complexity of the sign problem for the non-Gaussian representation is comparable to the case of Gaussian HS transformation with purely real exponents. Additionally, in non-Gaussian representation we can write the topological invariant for the detection of relevant saddle points, at least in the simplest case of one-site Hubbard model. 

We also describe some difficulties arising in QMC with continuous auxiliary fields due to the presence of the field configurations with zero fermionic determinant. First of all, the dimension of the manifolds formed by these field configurations is equal to $N-1$ in $\mathbb{R}^N$ if we work both in the limit with only complex or only real exponents. It means that in both limits we have the ``domain walls'' which divide the integration domain into disconnected regions. These ``domain walls'' are impenetrable for Hamiltonian updates of HMC algorithm thus it suffers from the ergodicity problems in both of these limits. This is an additional argument why the mixed regime with both types of exponents present in the action should be used. In the mixed regime the dimensionality of the manifolds with zero fermionic determinant is equal to $N-2$. 
The appearance of these configurations leads also to the heavy tailed distribution for  physical observables. Since the second moment is not  defined for these distributions, the statistical post-processing of the Monte Carlo data should be made carefully enough to give correct estimation of errors.

\section*{Acknowledgements}
We thank Prof. F.~Assaad and P.~Buividovich for interesting and motivating discussions. We also thank C.~Winterowd and D.~Smith for proofreading and providing useful comments. The work by M. U. was supported by the Deutsche Forschungsgemeinschaft (DFG), grant BU 2626/2-1.   
S.V. was supported by A. von Humboldt foundation (S. Kowalewskaja award to P. Buividovich).

%TODO: Provide arxiv preprint numbers for literature

\end{document}